\documentclass[final]{jfm}

\usepackage[colorlinks,citecolor=blue]{hyperref}
\usepackage{graphicx}
\usepackage{epstopdf, epsfig}

\usepackage{amsmath}
\shorttitle{Bursting Bubble in a Viscoplastic Medium}
\shortauthor{V. Sanjay, D. Lohse and M. Jalaal}
\title{Bursting Bubble in a Viscoplastic Medium}
\author{Vatsal Sanjay\aff{1}\corresp{\email{vatsalsanjay@gmail.com}},
 Detlef Lohse{\aff{1}$^,$\aff{2}\corresp{\email{d.lohse@utwente.nl}}}
	\and 	Maziyar Jalaal{\aff{3}$^,$\aff{4}}\corresp{\email{m.jalaal@uva.nl}}}
\affiliation{\aff{1}Physics of Fluids Group, Max Planck Center for Complex Fluid Dynamics,\\ MESA+ Institute and J.M. Burgers Center for Fluid Dynamics,\\
	University of Twente, P.O. Box 217, 7500 AE Enschede, the Netherlands
	\aff{2}Max Planck Institute for Dynamics and Self-Organisation, 37077 Göttingen, Germany
	\aff{3} Department of Applied Mathematics and Theoretical Physics, University of Cambridge, Cambridge CB3 0WA, United Kingdom
	\aff{4} Van der Waals–Zeeman Institute, Institute of Physics, University of Amsterdam, 1098XH Amsterdam, The Netherlands}

\usepackage{xcolor}

\begin{document}

\maketitle

\begin{abstract}
When a rising bubble in a Newtonian liquid reaches the liquid-air interface, it can burst, leading to the formation of capillary waves and a jet on the surface. Here, we numerically study this phenomenon in a yield stress fluid. We show how viscoplasticity controls the fate of these capillary waves and their interaction at the bottom of the cavity. Unlike Newtonian liquids, the free surface converges to a non-flat final equilibrium shape once the driving stresses inside the pool fall below the yield stress. Details of the dynamics, including the flow's energy budgets, are discussed. The work culminates in a regime map with four main regimes with different characteristic behaviours.
\end{abstract}

\section{Introduction}\label{Sec::introduction}
Bubble bursting processes abound in nature and technology and have been studied for long in fluid mechanics \citep{liger2008recent}. For example, they play a vital role in transporting aromatics from champagne \citep{liger2012physics,vignes2013fizzling,ghabache2014physics,ghabache2016evaporation}, and pathogens from contaminated water \citep{poulain2018biosurfactants,bourouiba2021fluid}. The process is also responsible for forming sea spray due to ejecting myriads of droplets \citep{macintyre1972flow, singh2019numerical}. Bursting bubbles also play an important role in geophysical phenomena such as volcanic eruptions \citep{gonnermann2007fluid}.

\begin{figure}
	\centerline{\includegraphics[width=0.67\linewidth]{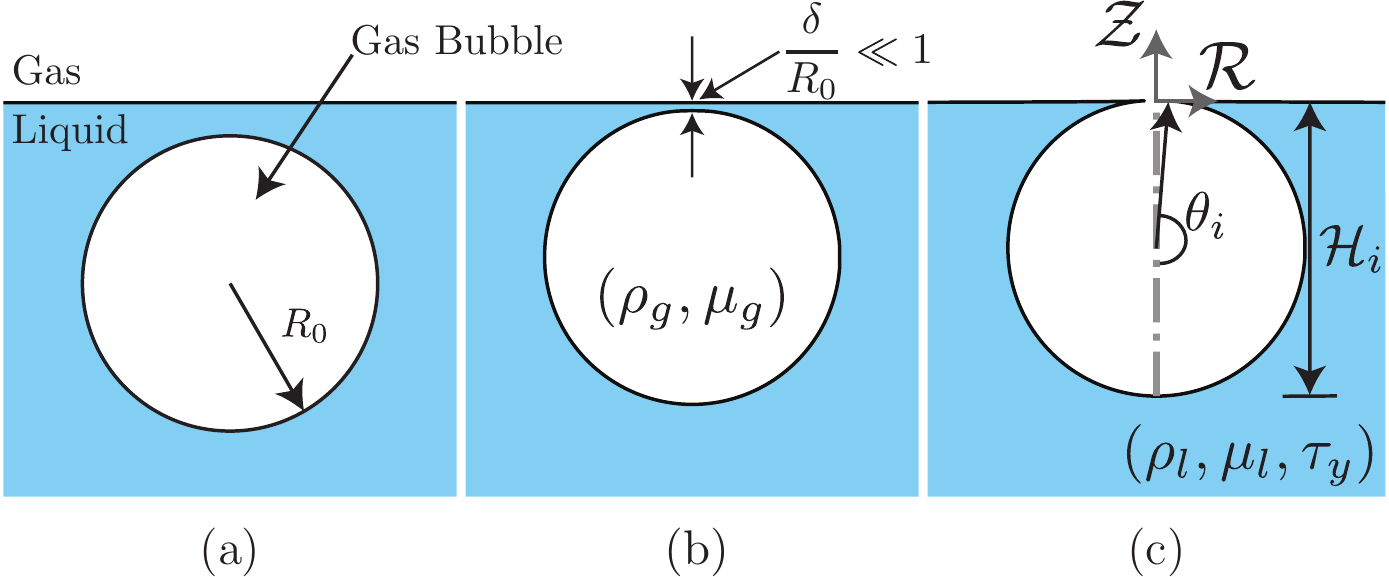}}
	\caption{Schematics for the process of a bursting bubble: (a) A gas bubble in bulk. (b) The bubble approaches the free surface forming a liquid film (thickness $\delta$) between itself and the free surface. (c) A bubble cavity forms when the thin liquid film disappears.}
	\label{fig:Schematic}
\end{figure}

In Newtonian liquids, the bubble bursting mechanism is controlled by buoyancy, surface tension, and viscosity. First, the air bubble (figure~\ref{fig:Schematic}(a)) being lighter than the surrounding medium, rises and approaches the liquid-air interface (figure~\ref{fig:Schematic}(b)). The thin film between the bubble and the free surface then gradually drains \citep{toba1959drop, princen1963shape} and eventually ruptures, resulting in an open cavity (figure~\ref{fig:Schematic}(c), \cite{mason1954bursting}). The collapse of this cavity leads to a series of rich dynamical processes that involve capillary waves  \citep{zeff2000singularity, duchemin2002jet} and may lead to the formation of Worthington jet \citep{gordillo2019capillary}. In some cases, the jet might break via a Rayleigh-Plateau instability, forming droplets \citep{ghabache2014physics,ghabache2016size}. The phenomenon is so robust that it even occurs in soft granular matter, when a rising bubble also bursts at the surface, leading to a granular jet \citep{lohse2004impact}.

The earlier work on bursting bubbles used Boundary Integral Methods in an inviscid limit \citep{boulton1993gas,longuet1995critical}. However, the progress in the Direct Numerical Simulation (DNS) tools for multiphase flows \citep{tryggvason2011direct, popinet2003gerris, popinet2009accurate} has resulted in models that consider the effects of viscosity. In fact, some recent studies revealed how liquid's viscosity affects the dynamics of the bursting bubbles \citep{deike2018dynamics, gordillo2019capillary}.

For Newtonian liquids, \cite{deike2018dynamics} have provided quantitative cross-validation of the numerical and experimental studies. They have also given a complete quantitative description of the influence of viscosity, gravity, and capillarity on the process, extending the earlier work of \cite{duchemin2002jet}. More recently, the experiments and simulations are complemented by theoretical frameworks  \citep{gordillo2019capillary, ganan2017revision}, resulting in a profound understanding of the physics of bubble bursting in Newtonian fluids. Appendix~\ref{App::Validation} provides more details on the previous studies in the Newtonian limit and compares our results with those available in the literature.

Notably, despite many applications, such as in the food industry and geophysics, the influence of rheological properties on the collapse of bubble cavities is yet to be understood. Here, we study the dynamics of bursting bubbles in a viscoplastic medium using Direct Numerical Simulations (DNS). Viscoplastic or yield stress fluids manifest a mix of solid and fluid behaviour. The materials behave more like an elastic solid below critical stress (yield stress); however, they flow like a viscous liquid above this critical stress. Readers can find detailed reviews on yield stress fluids in \cite{bird1983rheology,coussot2014yield,balmforth2014yielding,bonn2017yield}.

Previous experiments and simulations have been reported for trapped bubbles in a viscoplastic medium \citep{dubash2004conditions,sun2020dynamic, de2019oscillations}, rising bubbles in yield stress fluids \citep{sikorski2009motion,tripathi2015bubble, lopez2018rising,singh2008interacting, dimakopoulos2013steady, tsamopoulos2008steady, mougin2012significant}, and bubbles moving inside tubes filled with viscoplastic fluids \citep{jalaal2016long, laborie2017yield, zamankhan2018steady}. We will show that the introduction of non-Newtonian properties can significantly influence the bursting behaviour of bubbles on a free surface. At moderate values of yield stress, the collapse of the cavity can still lead to the formation of a Worthington jet, but the droplet formation might be suppressed. At high yield stress values, the unyielded region of the viscoplastic fluid can seize the collapse of this cavity, which leads to distinct final crater shapes.

The paper is organized as follows: \S~\ref{Sec::ProblemDescription} describes the problem and the governing parameters. \S~\ref{Sec::Phenomenological} provides a phenomenological analysis, and \S~\ref{Sec::Energy} presents the different modes of energy transfer during the viscoplastic bursting process. \S~\ref{Sec::EquilibriumStates} presents the final equilibrium shapes. The work culminates in \S~\ref{Sec::RegimeMaps} where we summarize the different regimes observed in the process of bursting in a phase diagram. The paper ends with conclusions in \S~\ref{Sec::Conclusion}.

\section{Numerical framework \& problem description}\label{Sec::ProblemDescription}
\subsection{Governing equations}\label{Sec::DimensionlessForm}
We consider the burst of a small axisymmetric bubble at a surface of an incompressible Bingham fluid. To nondimensionalise the governing equations, we remove the length and velocity scales using the initial bubble radius $R_0$ and inertia-capillary velocity $V_\gamma$, scales, respectively. Pressure and stresses are scaled with the characteristic capillary pressure $\tau_\gamma$ (see Appendix~\ref{App::GoverningEquations}). The dimensionless equations for mass and momentum conservation, for the liquid phase, then read

\begin{align}
\nabla \cdot \boldsymbol{u}&=0,\label{Eqn::FinalFormLiqContinuity}\\
	\frac{\partial\boldsymbol{u}}{\partial t} + \nabla\boldsymbol{\cdot}\left(\boldsymbol{uu}\right) &= -\nabla p + \nabla\boldsymbol{\cdot}\boldsymbol{\tau} - \mathcal{B}o\,\hat{\boldsymbol{e}}_{\boldsymbol{\mathcal{Z}}}\label{Eqn::FinalFormLiqNS},
\end{align}
where $\boldsymbol{u}$ is the velocity vector, $t$ is time, $p$ is the pressure and $\boldsymbol{\tau}$ represents the deviatoric stress tensor. We use a regularized Bingham model with
\begin{align}
    \boldsymbol{\tau} &= 2\,\text{min}\left(\frac{\mathcal{J}}{2\|\boldsymbol{\mathcal{D}}\|} + \mathcal{O}h, \mathcal{O}h_\text{max}\right)\boldsymbol{\mathcal{D}},\label{Eqn::RegularizedForm}
\end{align}
where $\|\boldsymbol{\mathcal{D}}\|$ is the second invariant of the deformation rate tensor, $\boldsymbol{\mathcal{D}}$, and $\mathcal{O}h_{\text{max}}$ is the viscous regularisation parameter. The three dimensionless numbers controlling the equations above are the plastocapillary number $\left(\mathcal{J}\right)$, which accounts for the competition between the capillary and yield stresses, the Ohnesorge number $\left(\mathcal{O}h\right)$ that compares the inertial-capillary to inertial-viscous time scales, and the Bond number $\left(\mathcal{B}o\right)$, which compares gravity and surface tension forces:
\begin{equation}\label{Eqn::DimensionlessNumbers}
	\mathcal{J} = \frac{\tau_yR_0}{\gamma},\,\,\mathcal{O}h = \frac{\mu_l}{\sqrt{\rho_l\gamma R_0}},\,\,\mathcal{B}o = \frac{\rho_l gR_o^2}{\gamma}.
\end{equation}

\noindent Here, $\gamma$ is the liquid-gas surface tension coefficient, and $\tau_y$ and $\rho_l$ are the liquid's yield stress and density, respectively. Next, $\mu_l$ is the constant viscosity in the Bingham model. Note that in our simulations, we also solve the fluid's motion in the gas phase, using a similar set of equations (see Appendix~\ref{App::GoverningEquations}). Hence, the further relevant non-dimensional groups in addition to those in equation~(\ref{Eqn::DimensionlessNumbers}) are the ratios of density $\left(\rho_r = \rho_g/\rho_l\right)$ and viscosity $\left(\mu_r = \mu_g/\mu_l\right)$. In the present study, these ratios are kept fixed at $10^{-3}$ and $2 \times 10^{-2}$, respectively. 

\subsection{Method}\label{Sec::Method}
For our calculations, we use the free software program Basilisk C \citep{basiliskPopinet, popinet2015quadtree}.  The code uses a Volume of Fluid (VoF) technique \citep{tryggvason2011direct} to track the interface, introducing a concentration field $c$, that satisfies the scalar advection equation. Hence, equations~(\ref{Eqn::FinalFormLiqContinuity}) -~(\ref{Eqn::FinalFormLiqNS}) and their counterparts for the gas phase are solved using the one-fluid approximation, where the surface tension acts as a body force, $\boldsymbol{f}_\gamma$ only at the gas-liquid interface \citep{brackbill1992continuum, popinet2009accurate}. In dimensionless form,
\begin{equation}\label{Eqn::SurfaceTension}
	\boldsymbol{f}_\gamma = \kappa\delta_s\hat{\boldsymbol{n}} \approx \kappa\boldsymbol{\nabla}c,
\end{equation}
where, $\delta_s$ is the interface Dirac function, $\hat{\boldsymbol{n}}$ is a unit vector normal to the interface \citep{tryggvason2011direct}, and $\kappa$ is the curvature of the interface, $z = S(r)$, given by \citet[p.~14-16]{deserno2004notes}:
\begin{equation}\label{Eqn::curvature}
	\kappa = \frac{\frac{d^2S}{dr^2}}{\left(1 + \left(\frac{dS}{dr}\right)^2\right)^{3/2}} + \frac{\frac{dS}{dr}}{r\left(1 + \left(\frac{dS}{dr}\right)^2\right)^{1/2}}.
\end{equation}

In Basilisk C, the curvature in equation~(\ref{Eqn::curvature}) is calculated using the height-function method.  As the surface-tension scheme is explicit in time, the maximum time-step is maintained at most at the oscillation period of the smallest wave-length capillary wave \citep{popinet2009accurate, basiliskPopinet2}.
Note that the curvature above is, in fact, the dimensionless capillary pressure. Hence, in the text, the wave with the largest curvature is called the \lq\lq strongest wave\rq\rq.

Basilisk C also provides Adaptive Mesh Refinement (AMR). We use this feature to minimize errors in the VoF tracer (tolerance threshold: $10^{-3}$) and interface curvature (tolerance threshold: $10^{-4}$). Additionally, we also refine based on velocity (tolerance threshold: $10^{-2}$) and vorticity (tolerance threshold: $10^{-3}$) fields to accurately resolve the regions of low strain rates. For AMR, we use a grid resolution such that the minimum cell size is $\Delta = R_0/512$, which implies that to get similar results, one will need $512$ cells across the bubble radius while using uniform grids. We have also carried out extensive grid independence studies to ensure that changing the grid size does not influence the results. Moreover, we employ free-slip and no-penetration boundary conditions for both liquid and gas at the domain boundaries.  For pressure, zero gradient condition is employed at the boundaries. For the cases where the Worthington jet breaks into droplets, an outflow boundary condition is employed at the top boundary to ensure that the drop does not bounce off the boundary. These boundaries are far away from the bubble (size of the domain is $8R_0$)  such that they do not influence the process. 

Note that our numerical method uses a regularized form of the Bingham constitutive equations (see equation~(\ref{Eqn::RegularizedForm}) and Appendix~\ref{App::GoverningEquations}). Hence, we cannot resolve the exact position of the yield surface as $\|\boldsymbol{\mathcal{D}}\|$ is never precisely zero. However, we can safely assume that low values of $\|\boldsymbol{\mathcal{D}}\|$ will be associated with the plastic regions. In our simulations, $\mathcal{O}h_{\text{max}}=10^8$. We have ensured that our results are independent of this regularisation parameter (Appendix~\ref{App::OhMax}). The regularisation of the constitutive model also forces us to choose a criterion for the stoppage time, $t_s$. In our simulations, we consider a significantly small cut-off kinetic energy to stop the simulations (see Appendix~\ref{App::StoppageTime} for details). 

\subsection{Initial Condition}\label{Sec::Initial}
This initial shape of a bubble at a fluid-fluid interface (figure~\ref{fig:Schematic}(b)) can be calculated by solving the Young-Laplace equations to find the quasi-static equilibrium state for an arbitrary Bond number, $\mathcal{B}o$ (see \citet{lhuissier2012bursting, walls2015jet, deike2018dynamics,magnaudet2020particles}). As a starting point, this study only concerns the limit of $Bo \rightarrow 0$, i.e., when capillary effects dominate the gravitational ones. We choose $\mathcal{B}o = 10^{-3}$ for all the simulations in this work.  For this value, the initial bubble is nearly spherical in a surrounding Newtonian liquid.  Note that the bubble sphericity is a crucial assumption (simplification) in our work. The actual initial shape of the bubble depends on its size ($\mathcal{B}o$), material properties ($\mathcal{J}$, $\mathcal{O}h$), the method of generation, and its dynamics before approaching the interface \citep{dubash2004conditions, dimakopoulos2013steady}. Furthermore, for a bubble to rise in a viscoplastic medium, the buoyancy forces should be strong enough to yield the flow \citep{dubash2004conditions, sikorski2009motion, balmforth2014yielding}, \textit{i.e.,} $\mathcal{B}o \gg \mathcal{J}$. Hence, non-spherical and non-trivial shapes could be expected \citep{lopez2018rising}. For such a limit, one should first solve the full dynamics of rising bubbles to achieve the correct initial condition for the bursting problem. Note that low $\mathcal{B}o$ bubbles could still form near a free-surface in other situations. One example is the process of Laser-Induced Forward Transfer (LIFT), in which a laser pulse generates a bubble near the free surface of a viscoplastic liquid \citep{jalaal2019laser}.

For our given initial shape, the value of $\mathcal{J}$ varies between 0 and 64. This range is selected such that we will study a full range of yield stress effects, from the Newtonian limit ($\mathcal{J}=0$) to a medium that barely deforms due to a large yield stress ($\mathcal{J}=64$).

Following the common assumption in these types of problems \citep{deike2018dynamics, gordillo2019capillary}, we assume that the thin liquid cap of thickness $\delta$ (figure~\ref{fig:Schematic}(b)) disappears at $t = 0$, resulting in the configuration shown in figure~\ref{fig:Schematic}(c), i.e. the initial condition for our simulations. In figure~\ref{fig:Schematic}(c), $\left(\mathcal{R}, \mathcal{Z}\right)$ denotes the radial and axial coordinate system. Furthermore, $\mathcal{H}_i \approx 2$ is the initial bubble depth, and $\theta_i$ is the initial location of the cavity-free surface intersection. Note that the curvature diverges at this intersection in such a configuration. We smooth the sharp edge using a small circular arc to circumvent this singularity, introducing a rim with a finite curvature $\kappa_0$ that connects the bubble to the free surface. We ensured that the curvature of the rim is high enough such that the subsequent dynamics are independent of its finite value (for details, see Appendix~\ref{App::FilletCurvature}).

\section{Effects of yield stress on the bursting bubble}\label{Sec::Phenomenological}
\subsection{Phenomenology}\label{Sec::Phenomenological1}
This section describes the dynamics of bursting bubbles and the qualitative effects of the plastocapillary number $\left(\mathcal{J}\right)$. Figure~\ref{fig:J_Variation} illustrates four representative cases for this purpose (videos are available in the supplementary material). For a Newtonian liquid (figure~\ref{fig:J_Variation}(a), $\mathcal{J} = 0$), the retraction of the rim leads to the formation of capillary waves.

Part of these waves travels away from the cavity, forming regions of small strain rates (black dots in figure~\ref{fig:J_Variation}(a): $t = 0.45$), which are advected with the train of capillary waves. 
Meanwhile the other part of the waves travel down the cavity (figure~\ref{fig:J_Variation}(a): $t = 0.1$) and focuses on the cavity's bottom (figure~\ref{fig:J_Variation}(a): $t = 0.45$). 

Consequently, a Worthington jet is formed as depicted in figure~\ref{fig:J_Variation}(a): $t = 0.65$. Furthermore, due to the conservation of momentum, a high-velocity jet is also formed opposite to this Worthington jet inside the liquid pool (figure~\ref{fig:J_Variation}(a): $t = 0.65$). The Worthington jet can then break into multiple droplets due to the Rayleigh-Plateau instability \citep{walls2015jet}. In the Newtonian limit, the flow continues until the free surface is fully flat, when the surface energy is minimized (figure~\ref{fig:J_Variation}(a): $t = 4.00$).
 \begin{figure}
	\centerline{\includegraphics[width=\linewidth]{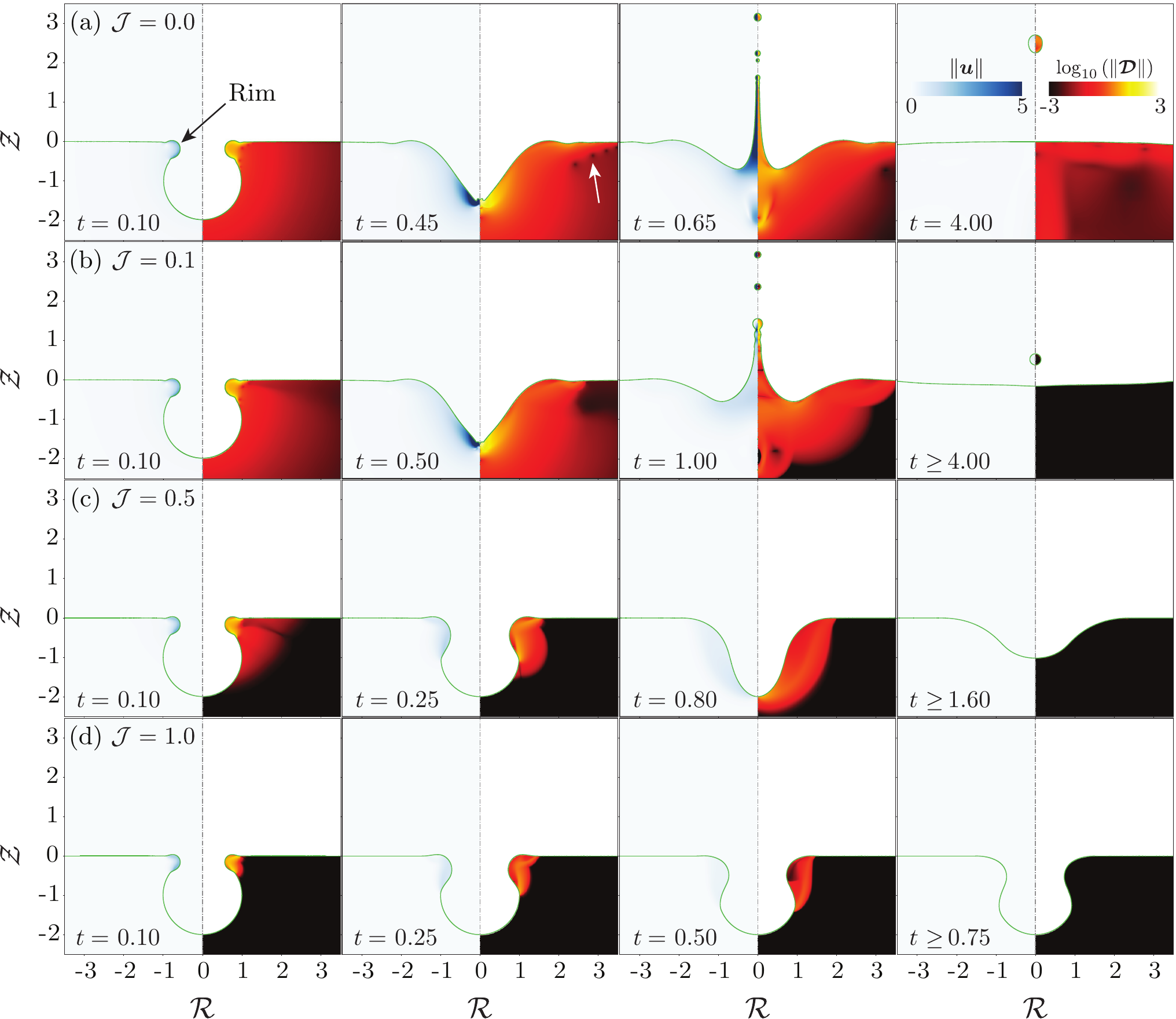}}
	\caption{Bursting bubble dynamics for different plastocapillary numbers. (a) $\mathcal{J} = 0.0$: A typical case with a Newtonian liquid medium, (b) $\mathcal{J} =0.1$: A weakly viscoplastic liquid medium in which the process still shows all the major characteristics of the Newtonian liquid, (c) $\mathcal{J} = 0.5$: A case of moderate yield stress whereby the jetting is suppressed, nonetheless the entire cavity still yields, and (d) $\mathcal{J} = 1.0$: A highly viscoplastic liquid medium whereby a part of the cavity never yields. The left part of each panel shows the magnitude of the velocity field, and the right part shows the magnitude of the deformation tensor on a $\log_{10}$ scale. The transition to the black region (low strain rates) marks the yield-surface location in the present study. The time instances in this figure are chosen to show significant events throughout the process of bursting bubbles for different $\mathcal{J}$ numbers. For all the cases in this figure, $\mathcal{O}h = 10^{-2}$. Videos (S1 - S4) are available in the supplementary material.}
	\label{fig:J_Variation}
\end{figure}

The introduction of the yield stress, in general, slows down the flow due to a larger apparent viscosity. Remarkably, even at large yield stresses, the early time dynamics near the retracting rim remain unchanged due to the highly curved interface, as clearly shown in the first panels ($t = 0.1$) of figure~\ref{fig:J_Variation} (a-d).  
On the contrary, the anatomy of the flow inside the pool is considerably affected due to the yield stress. At low yield stresses ($\mathcal{J} = 0.1$ in figure~\ref{fig:J_Variation}(b): $t = 0.1$), everywhere near the bubble cavity yields at early times. However, as the values of plastocapillary number increases, the size of the yielded region decreases ($\mathcal{J} = 0.5\, \&\,\mathcal{J} = 1.0$ in figures~\ref{fig:J_Variation}(c) and~\ref{fig:J_Variation}(d), respectively). 

Furthermore, at low values of $\mathcal{J}$, the flow focusing at the bottom of the cavity persists (figure~\ref{fig:J_Variation}(b): $t = 0.50$), although, due to the increased dissipation, it is less vigorous. As a result, the jet formed post-collapse is thicker, slower, and less prominent (figure~\ref{fig:J_Variation}(b): $t = 1.00$) as compared to the Newtonian case (figure~\ref{fig:J_Variation}(a): $t = 0.65$). Notably, for small values of $\mathcal{J}$, the Worthington jet still forms and breaks up into droplets due to the Rayleigh-Plateau instability. 

Note that unlike for the Newtonian case, where the final shape is always a flat free-surface, a viscoplastic medium (i.e., finite $\mathcal{J}$) comes to a halt when stress inside the liquid drops below the yield stress. Hence, the final state can feature non-zero surface energy (figure~\ref{fig:J_Variation}(b): $t \ge 4.00$). 

At higher values of $\mathcal{J}$, the capillary waves are so damped that flow focusing at the bottom of the cavity vanishes. At moderate $\mathcal{J}$ numbers (figure~\ref{fig:J_Variation}(c) where $\mathcal{J} = 0.5$), the capillary waves are still strong enough to travel over the entire cavity (figure~\ref{fig:J_Variation}(c): $t = 0.25 - 0.80$). As a result, the entire cavity yields, nonetheless,  the final shape still features a deep crater (figure~\ref{fig:J_Variation}(c): $t = 1.60$). 
On further increasing $\mathcal{J}$ such that yield stress is as strong as the capillary stress $\left(\tau_y \sim \gamma/R_0\text{, i.e.,}\,\mathcal{J} \sim \mathcal{O}\left(1\right)\right)$, the capillary waves do not yield the entire cavity (figure~\ref{fig:J_Variation}(c): $t = 0.25 - 0.75$). Hence, the final shape furnishes a deep crater that stores a large surface energy, contrary to the final shapes at small $\mathcal{J}$  values.

In this section, we mainly focus on the effect of yield stress (via $\mathcal{J}$) on the process of bursting bubbles. Appendix~\ref{App::OhsVariation} contains the discussion on the effect of $\mathcal{O}h$. Furthermore, in the subsequent sections, we will discuss the features explained above in more quantitative details. \S~\ref{Sec::CapillaryWaves} and \S~\ref{Sec::JetFormation} then delineate the travelling capillary waves and the subsequent jet formation (or lack of it), respectively.  

\subsection{Capillary waves in the presence of yield stress}\label{Sec::CapillaryWaves}
\begin{figure}
	\centerline{\includegraphics[width=\linewidth]{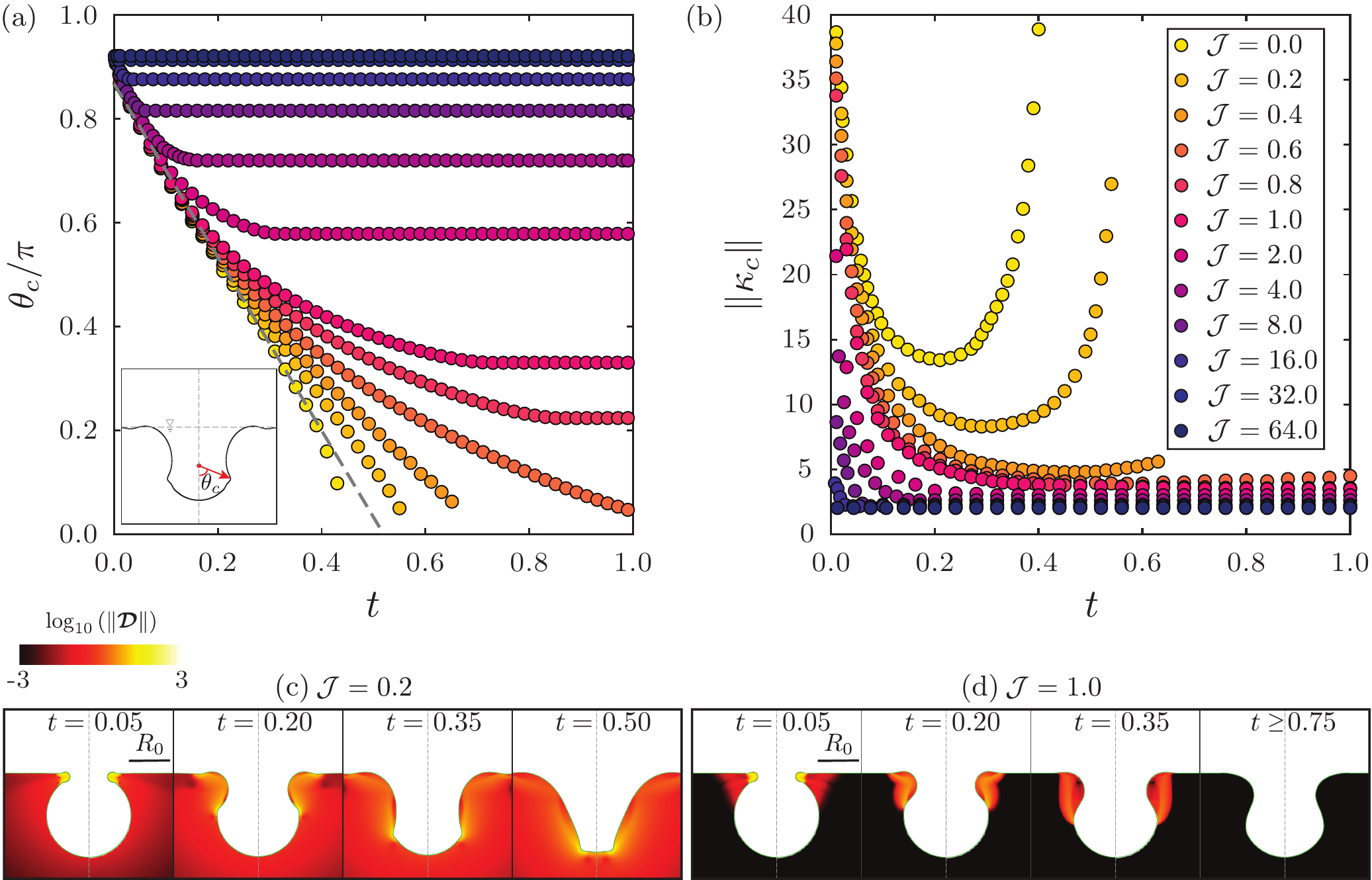}}
	\caption{Effects of viscoplasticity on the traveling capillary waves: (a) Variation of the location $\left(\theta_c\right)$ of strongest capillary with time. The gray dotted line denotes the Newtonian limit, $\theta_c -\theta_i \sim -V_\gamma\,t$ as described by \citet{gordillo2019capillary}. (b) Variation of the strength $\left(\|\kappa_c\|\right)$ of the strongest capillary wave with time. Snapshots of the deformation tensor modulus $\|\boldsymbol{\mathcal{D}}\|$ for (c) $\mathcal{J} =$ 0.2, and (d) $\mathcal{J} =$ 1.0. For all the cases in this figure, $\mathcal{O}h = 10^{-2}$. Videos (S5 - S7) are available in the supplementary material.}
	\label{fig:Jcar_EarlyTimes}
\end{figure}
Capillary waves are critical in the bubble bursting process \citep{gordillo2019capillary}. Initially, the breakage of the film and the retraction of the rim create a train of capillary waves of varying strengths \citep{gekle2009high}. However, sharper waves experience very high viscous dampening. As a result, the wave focusing and jet formation are controlled by the strongest wave, which is not ceased by viscous damping. We follow \citet{gordillo2019capillary} and track the strongest wave by chasing the maximum curvature of the free surface wave ($\|\kappa_c\|$). The location of this wave, on the cavity, is denoted by the angular position, $\theta_c$ (see inset in figure~\ref{fig:Jcar_EarlyTimes}(a)). 

For a Newtonian liquid $\left(\mathcal{J} = 0\right)$, at low Ohnesorge numbers (e.g., figure~\ref{fig:Jcar_EarlyTimes}), the strongest capillary wave propagates at a constant velocity $V_\gamma$, dashed line in figure~\ref{fig:Jcar_EarlyTimes}(a)). The viscous stress attenuates these waves but does not influence $\theta_c$. Previous studies \citep{krishnan2017scaling, gordillo2019capillary} have found similar results for Newtonian liquids (see Appendix~\ref{App::Validation} for more details). The strength of this wave decreases as it propagates down the cavity due to continuous viscous dissipation. Around $\theta_c \approx \pi/2$, the geometry changes leading to flow focusing resulting in an increase in the strength $\left(\kappa_c\right)$ of the wave (see figure~\ref{fig:Jcar_EarlyTimes}(c): $t = 0.2$ to $t = 0.35$). This minimum value of $\kappa_c$ non-linearly depends on $\mathcal{O}h$ (see \citet{gordillo2019capillary} \& Appendix~\ref{App::Validation} for details).

As shown in figure~\ref{fig:Jcar_EarlyTimes}(a) and~\ref{fig:Jcar_EarlyTimes}(b) (and also discussed in \S~\ref{Sec::Phenomenological1}), the initial changes in $\|\theta_c\|$ and $\|\kappa_c\|$ remain similar to the Newtonian limit, since the highly curved region near the initial rim retraction fully yields the fluid around it. As the flow develops, the plasticity effects become more pronounced as compared to the capillary effects, and the capillary waves no longer follow the path taken by their Newtonian counterpart.
The larger the value of $\mathcal{J}$, the sooner the dynamics of the capillary waves deviate from the Newtonian limit, and they become weaker. Eventually, the waves stop at a finite stoppage time, furnishing a finite final $\theta_c$ and $\|k_c\|$ (represented by $\theta_f$ and $\|\kappa_f\|$, respectively). In section \S~\ref{Sec::EquilibriumStates}, we will discuss the variation of these parameters for the final crater shapes.

\subsection{Jet formation in the presence of yield stress}\label{Sec::JetFormation}
 \begin{figure}
	\centerline{\includegraphics[width=\linewidth]{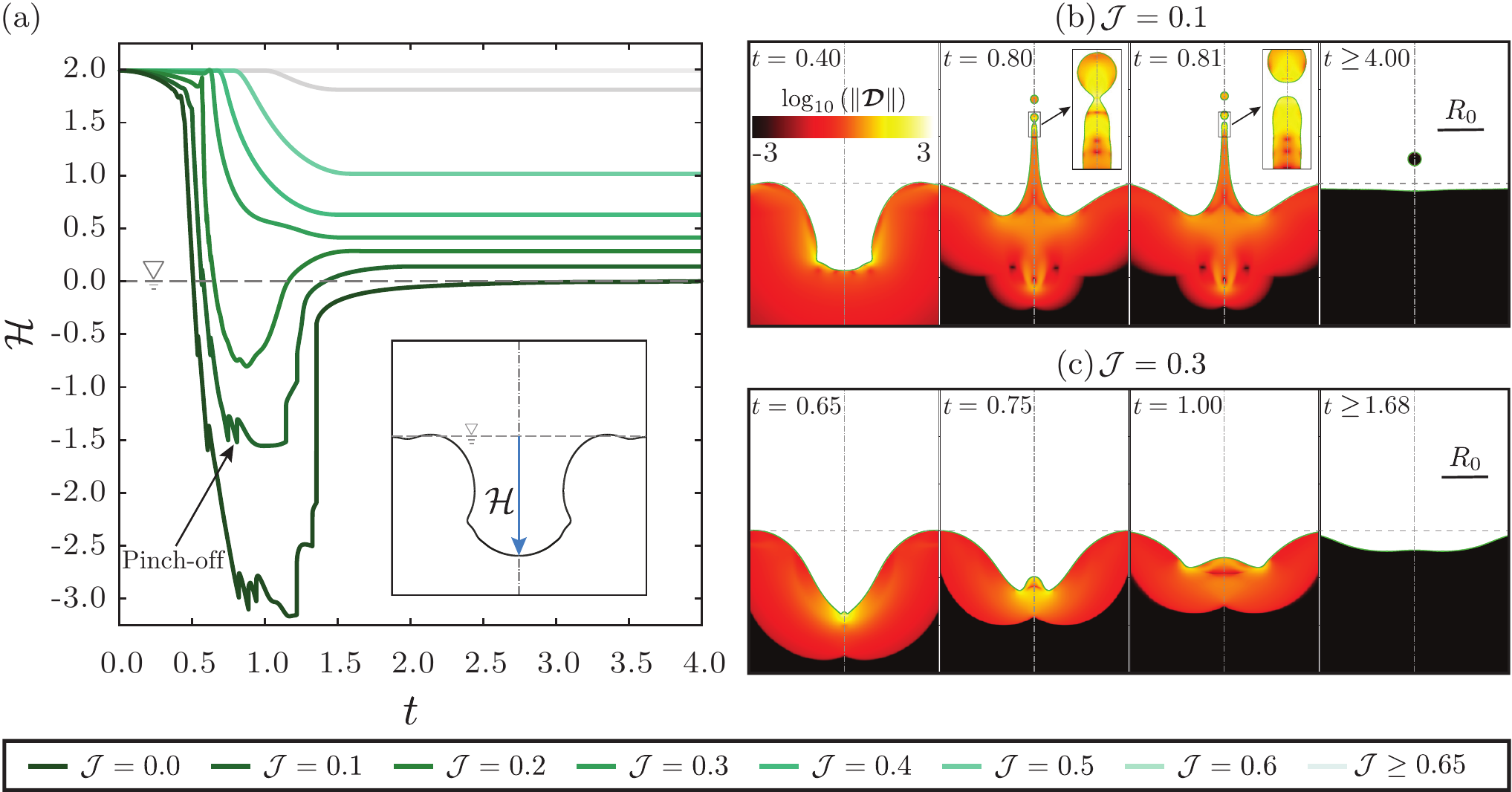}}
	\caption{Effects of viscoplasticity on the formation of the jet as a result of the collapsing cavity: (a) Variation of the depth $\mathcal{H}$ of the cavity at its axis with time. The inset shows the definition of $\mathcal{H}$. Modulus of the deformation tensor $\|\boldsymbol{\mathcal{D}}\|$ for the collapse of the bubble cavity and formation of the jet for (b) $\mathcal{J} =$  0.1 and (c) $\mathcal{J} =$ 0.3. Note that each kink in panel (a) is associated with the formation of a drop, as illustrated in the insets of the panel (b). For all the cases in this figure, $\mathcal{O}h = 10^{-2}$. Videos (S8 - S10) are available in the supplementary material.}
	\label{fig:Jcar_LateTimes}
\end{figure}
Another interesting feature of the bubble bursting process is the Worthington jet's formation as the bubble cavity collapses. To characterize this jet, we track the location $\left(\mathcal{H}\right)$ of the interface at the center $\mathcal{R} = 0$. 
In figure~\ref{fig:Jcar_LateTimes}(a) shows the temporal variation of $\mathcal{H}$ for different values of $\mathcal{J}$ at fixed $\mathcal{O}h = 10^{-2}$. As the waves propagate, the cavity begins to collapse; hence the value of $\mathcal{H}$ decreases. A jet forms when the bottom of the cavity crosses the free surface, and $\mathcal{H}$ becomes negative (see figure~\ref{fig:Jcar_LateTimes}(b)). For small values of $\mathcal{J}$, the Rayleigh-Plateau instability and a subsequent pinch-off occur, resulting in the kinks shown in figure~\ref{fig:Jcar_LateTimes}(a). The jets eventually retract, and $\mathcal{H}$ approaches $0$, i.e., a flat final interface. As the value of $\mathcal{J}$ increases, the final value of $\mathcal{H}$ increases, approaching the upper bound of $\mathcal{H} = \mathcal{H}_i = 2$, which is set by the initial condition (twice the bubble radius, i.e., the bottom of the cavity never yields). 
In fact, for $\mathcal{J} \ge 0.65$, this value remains unchanged, meaning the plug region attached to the bottom of the cavity never yields. Note that, in an intermediate range of $\mathcal{J} \sim 0.35$, the interplay of the capillary waves and the yield stress results in a dimple (underdeveloped jet) that never crosses the free surface (see figure~\ref{fig:Jcar_LateTimes}(c)). 

\section{What happens to the initial surface energy?}\label{Sec::Energy}
\begin{figure}
	\centerline{\includegraphics[width=\linewidth]{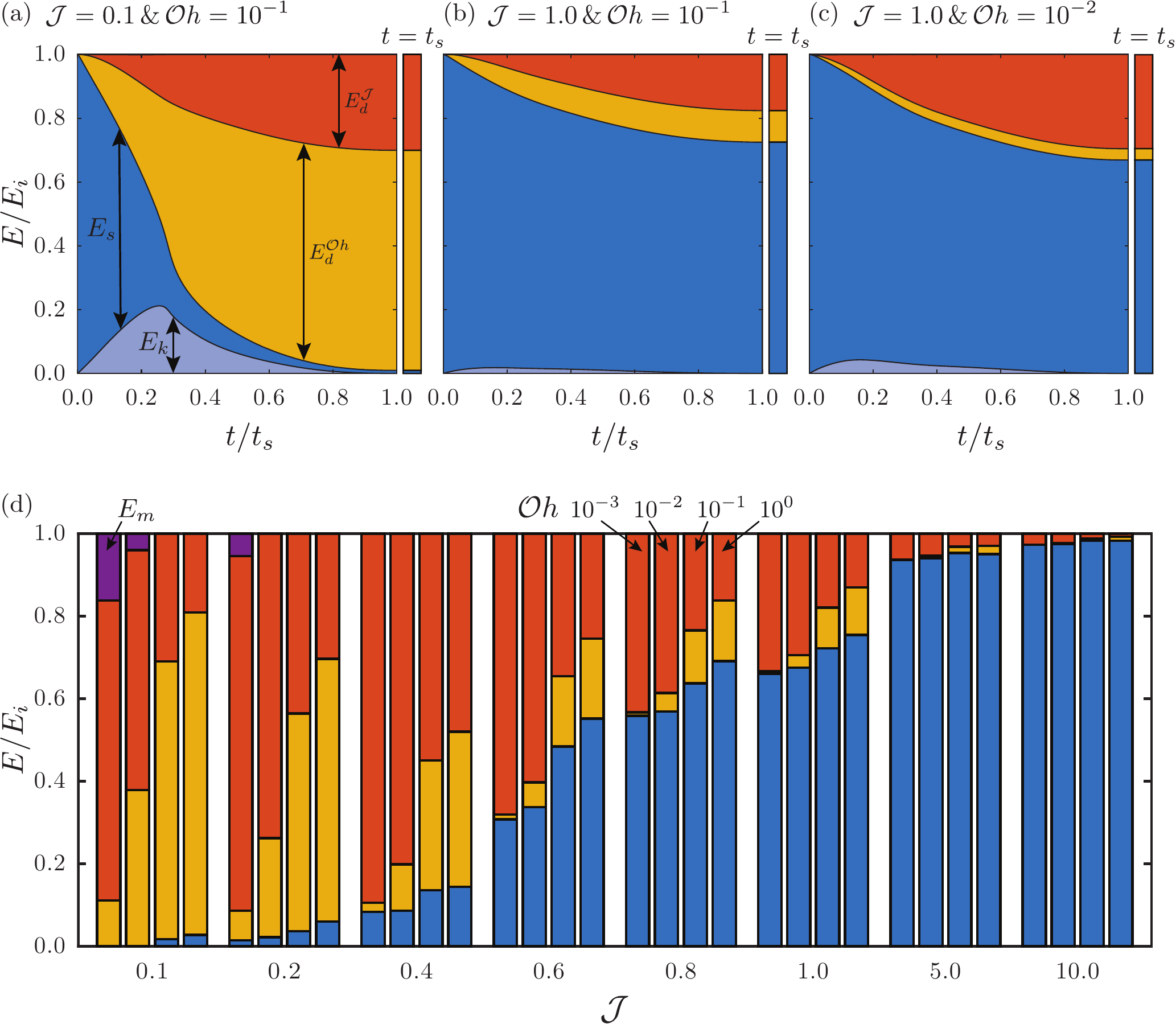}}
	\caption{Energy budget for the process of the bubble bursting in a viscoplastic medium: Temporal evolution of the different modes of energy transfers for (a) $\mathcal{J} =0.1$ \& $\mathcal{O}h =10^{-1}$, (b) $\mathcal{J} = 1.0$ \& $\mathcal{O}h =10^{-1}$, and (c) $\mathcal{J} =1.0$ \& $\mathcal{O}h =10^{-2}$. (d) Comparison of the energy footprint at the stoppage time, $t = t_s$ for different $\mathcal{J}$ and $\mathcal{O}h$.}
	\label{fig:EnergyBudget}
\end{figure}
To better understand the bubble bursting dynamics in a viscoplastic medium, we also looked at the energy budgets. The total energy $E$ is the sum of the total kinetic energy of the liquid pool $E_k$, its surface energy $E_s$ and the energy dissipation $E_d$. The latter contains two parts due to viscous ($E_d^{\mathcal{O}h}$) and yield stress ($E_d^\mathcal{J}$) contributions, $E_d = E_d^{\mathcal{O}h} +E_d^\mathcal{J}$. Lastly, small energies associated with jet breakup and airflow are summarized in $E_m$. Hence,
\begin{equation}\label{Eqn::ETotal}
E = E_k(t) + E_s(t) + E_d(t) + E_m(t) = E_i,
\end{equation}
where, $E_i$ is the initial energy that is purely the surface energy. Readers are referred to Appendix~\ref{App::EnergyBudgetApp} for details for calculating the energy budget. 

Figure~\ref{fig:EnergyBudget} shows three representative examples of these energy budgets, normalized by the initial energy $E_i$. In figure~\ref{fig:EnergyBudget}, the time is normalized by the stoppage time $t_s$. Panel a shows the temporal evolution of different modes of the energy transfer for a low $\mathcal{J}$ number. 
Initially, at $t = 0$, the system's total energy is stored as the bubble cavity's surface energy. 
As the flow starts, a part of this surface energy converts to the kinetic energy of the flow generated by the travelling capillary waves. The kinetic energy reaches a maximum when the capillary waves focus at the bottom of the cavity, as the focusing process forms a region of high velocity. At the instant of focusing, for $\mathcal{J} = 0.1\,\&\,\mathcal{O}h = 10^{-1}$, $\sim$ 60\% of the initial energy is still present in the system as a sum of the kinetic and surface energy of the liquid pool. Subsequently, the Worthington jet forms and high dissipation is observed due to an increase in the strain rate (see equation~(\ref{Eqn::Ed})). The surface energy decreases monotonically throughout the process and reaches a finite near-zero value at the stoppage time $t_s$. This behaviour is different from a Newtonian liquid, where the surface energy would become exactly zero, as $t \to \infty$.

For $\mathcal{J} = 1.0$ (figure~\ref{fig:EnergyBudget}(b) and~\ref{fig:EnergyBudget}(c)), initially, the surface energy decreases monotonically until it reaches a plateau at $t = t_s$. However, contrary to the example with a small value of $\mathcal{J}$, in these cases, a major part of the cavity never yields. Consequently, more than 70\% for $\mathcal{O}h =10^{-1}$ and over 60\% for $\mathcal{O}h =10^{-2}$ of the initial energy is still stored as the crater's surface energy. Also note that in the limit of large $\mathcal{J}$ (and low $\mathcal{O}h$), yield stress is responsible for the majority of energy dissipation, i.e., $E_d^{\mathcal{O}h} \ll E_d^\mathcal{J}$.

The energy footprint at $t = t_s$ gives insight into the bubble cavity's final static shape. Therefore, we compare these energies for different $\mathcal{J}$ and $\mathcal{O}h$ numbers in figure~\ref{fig:EnergyBudget}(d). For all the conditions, $E_k \to 0$, and only surface energy remain in the system at the stoppage time. Rest of the energy $\left(E_i - E_s\right)$ features as dissipation (except for those cases where drops form and $E_m$ is not negligible). 

For low values of $\mathcal{J}$, the final surface energy $E_s$ is close to zero because the final craters are shallow (see \S~\ref{Sec::EquilibriumStates} for details). This residual surface energy increases with increasing $\mathcal{J}$ and $\mathcal{O}h$; however, the dependency on $\mathcal{O}h$ is negligible at higher values of $\mathcal{J}$. Lastly, the dissipation due to the yield stress $\left(E_d^\mathcal{J}\right)$ contributes more to the overall dissipation for small $\mathcal{O}h$ numbers.

\section{Final crater shapes}\label{Sec::EquilibriumStates}
 \begin{figure}
	\centerline{\includegraphics[width=\linewidth]{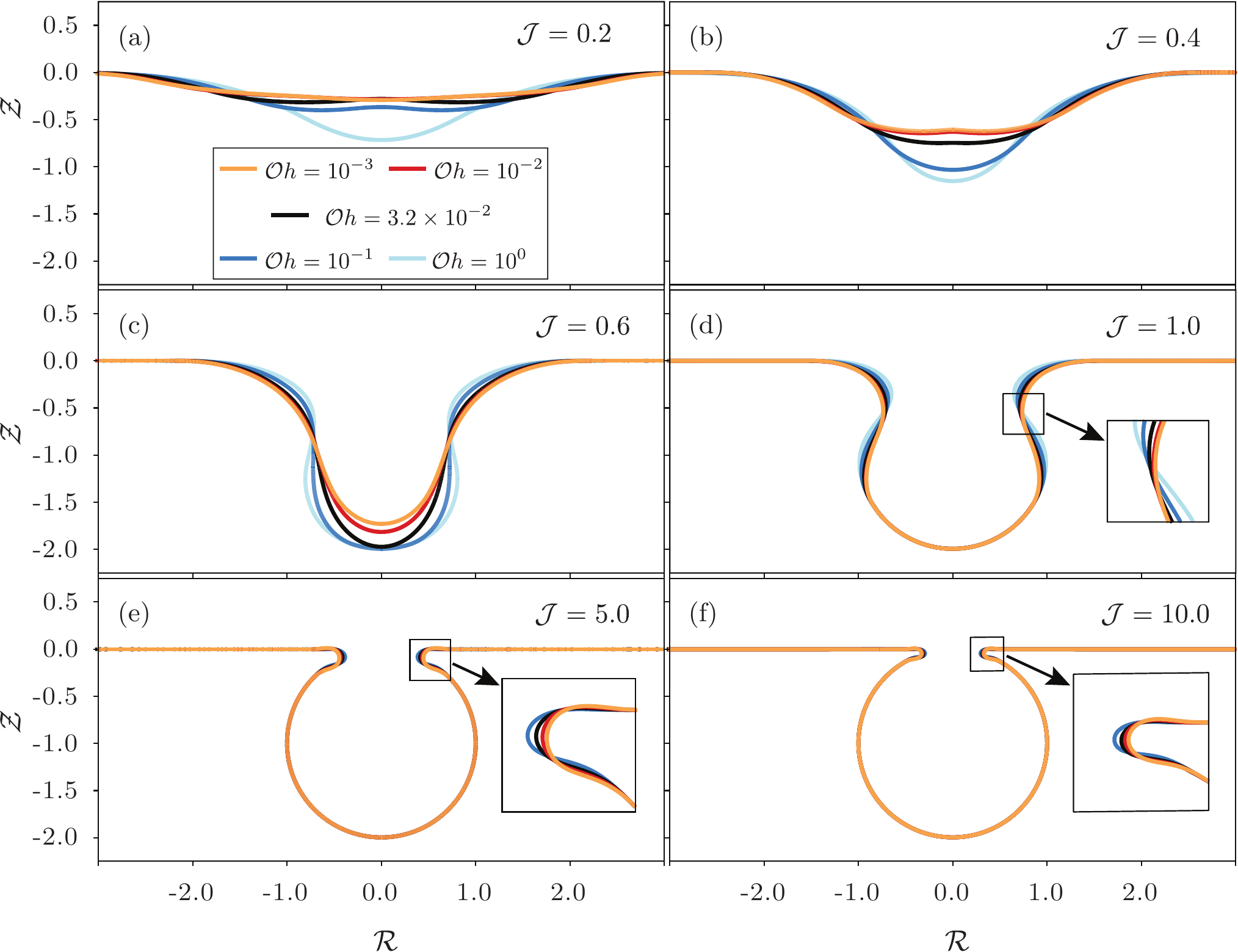}}
	\caption{Final crater shapes: Variation of the final shapes with the $\mathcal{O}h$ at (a) $\mathcal{J} =$ 0.2, (b) $\mathcal{J} =$ 0.4, (c) $\mathcal{J} =$ 0.6, (d) $\mathcal{J} =$ 1.0, (e) $\mathcal{J} =$ 5.0, and (f) $\mathcal{J} =$ 10.0.}
	\label{fig:Zoo_of_Final_Shapes}
\end{figure}
\begin{figure}
	\centerline{\includegraphics[width=\linewidth]{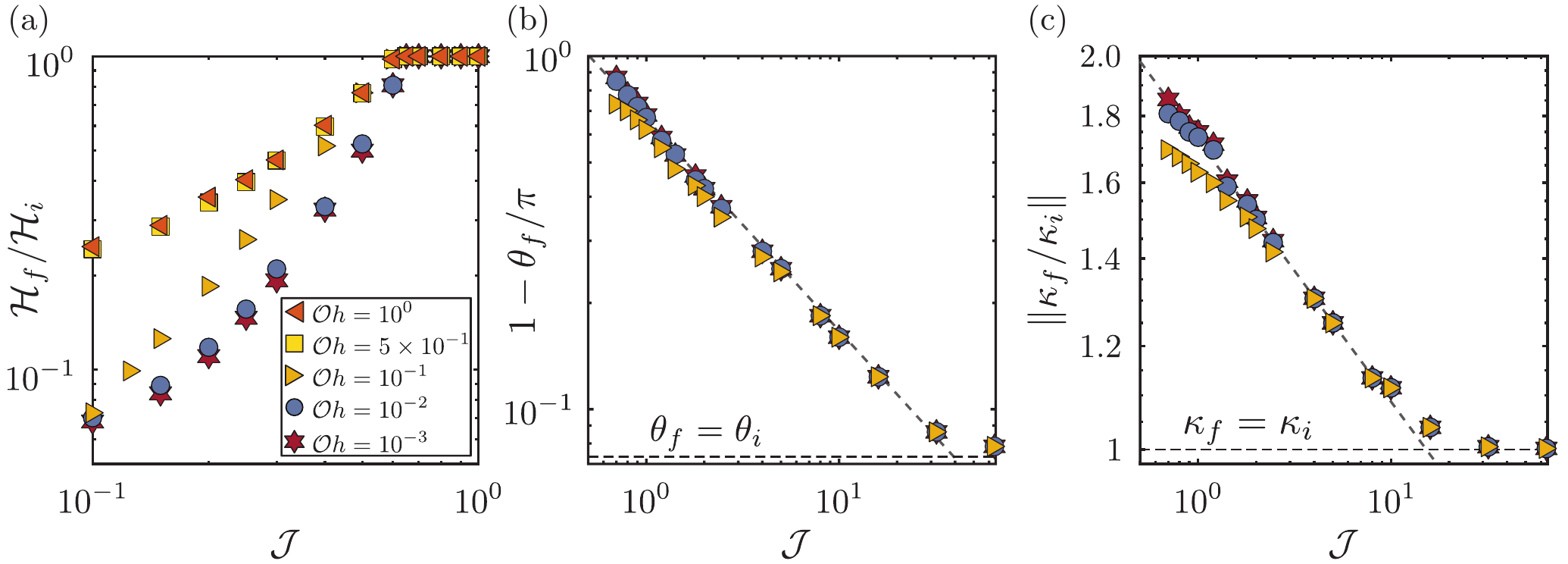}}
	\caption{Quantifying the characteristics of the final shapes as a function of $\mathcal{J}$ at different $\mathcal{O}h$: (a) Depth $\mathcal{H}_f$ of centre-line of the final cavity surface. (b) Location $\theta_f$ and (c) Strength $\|\kappa_f\|$ of the strongest capillary wave in the final crater. The grey dashed lines in panels b and c are guides to the eye.}
	\label{fig:Quantification_Final_Shapes}
\end{figure}
The process of bubble bursting in yield-stress fluids results in non-flat final shapes.
Figure~\ref{fig:Zoo_of_Final_Shapes} shows the final crater shapes as observed for different $\mathcal{J}$ and $\mathcal{O}h$ numbers, and figure~\ref{fig:Quantification_Final_Shapes} quantifies the different features of these final shapes by analysing the location $\left(\theta_f\right)$ and strength $\left(\|\kappa_f\|\right)$ of the strongest capillary wave, and the final depth of the crater $\left(\mathcal{H}_f\right)$. For the convenience of comparison, we normalize $\mathcal{H}_f$ by its initial value $\mathcal{H}_i \approx 2$, and $\|\kappa_f\|$ by the initial curvature of the cavity's bottom $\|\kappa_i\| \approx 2$. 

At low values of yield stress ($\mathcal{J} \le 0.4$), the final shape of the crater strongly depends on $\mathcal{O}h$ (see figures~\ref{fig:Zoo_of_Final_Shapes}(a-b), and~\ref{fig:Quantification_Final_Shapes}(a)).
When both $\mathcal{J}$ \& $\mathcal{O}h$ are small, the Worthington jet forms (see \S~\ref{Sec::JetFormation} for detailed discussions) which relaxes back towards the flat surface as $t \to \infty$. This jet relaxation results in shallow final cavities.
As $\mathcal{O}h$ increases, the viscose dissipation dominates the flow, the capillary waves are damped, and the change in the final cavity height becomes minimal; hence, $\mathcal{H}_f \sim \mathcal{H}_i$. In fact, for $\mathcal{O}h > 10^{-1}$, the capillary wave's amplitude is so close to zero that it becomes impossible to track.

As the value of yield stress ($\mathcal{J}$) increases, the effective viscosity of the domain increases and hence the initial cavity deforms less. For a highly plastic medium, the capillary waves cannot yield the entire cavity. As a result, $\mathcal{H}_f = \mathcal{H}_i$ for $\mathcal{J} \ge 0.65$, independent of the values of $\mathcal{O}h$. For higher $\mathcal{O}h$, this transition is reached (marginally) earlier (e.g., $\mathcal{J} \ge 0.5$ for $\mathcal{O}h \ge 1.0$).

For the cases where the bottom of the cavity never yields (figure~\ref{fig:Zoo_of_Final_Shapes}(d-f)), we characterize the final crater shapes based on the location and strength of the frozen capillary wave ($\theta_f$ and $\kappa_f$, respectively). The variations of these values are shown in figure~\ref{fig:Quantification_Final_Shapes}(b) and (c). As $\mathcal{J}$ increases, the final location of the wave is closer to the initial value, $\theta_f \approx \theta_i$. Similarly, the strength of the final wave approaches the value defined by the initial condition, as $\mathcal{J}$ increases. For this regime, the effects of $\mathcal{O}h$ on the final shape seem to be negligible for $\mathcal{J} > 2$. 

\section{Regime map}\label{Sec::RegimeMaps}
\begin{figure}
	\centerline{\includegraphics[width=0.5\linewidth]{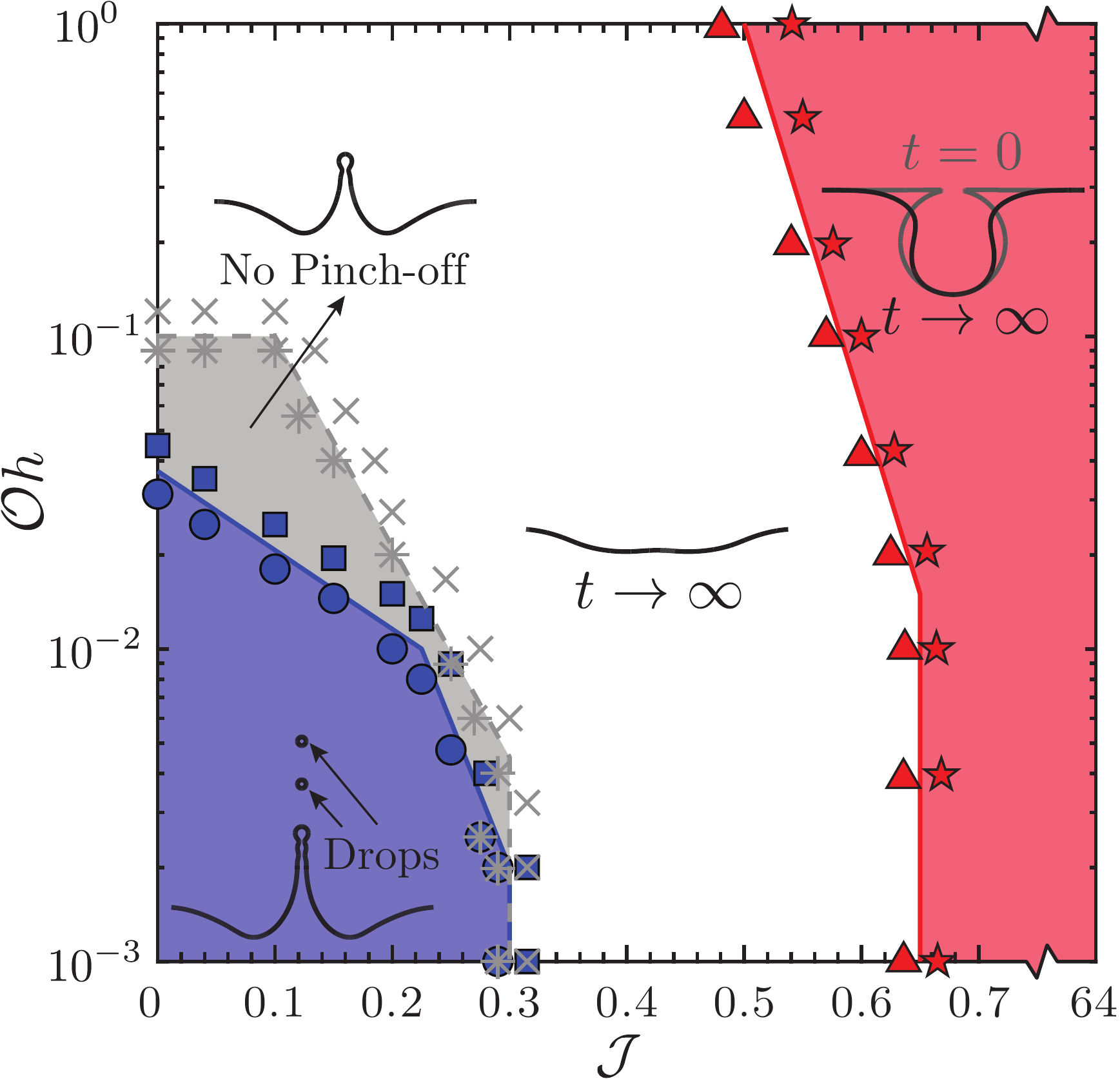}}
	\caption{Regime map in terms of the plastocapillary number $\mathcal{J}$ and the Ohnesorge number $\mathcal{O}h$ showing the transitions between the different categories identified in the current study. The insets show a representative case from each of the four regimes, namely formation of jet which breaks into droplets (blue), formation of jet without droplets (gray), the entire cavity collapses but the cavity center never crosses the initial pool free surface (white), and a part of the cavity never yields (red). The symbols represent simulations at the different transition lines.}
	\label{fig:Regime_Map}
\end{figure}

In the present study, the two crucial control parameters to describe the process of bursting bubbles in a viscoplastic medium are the plastocapillary number $\mathcal{J}$, and the Ohnesorge number $\mathcal{O}h$. This section uses these dimensionless numbers to summarize the observed features explained in the text, providing a regime map (or phase diagram). Figure~\ref{fig:Regime_Map} shows this map for the bursting bubble process in a viscoplastic medium. Note that we have run more than 750 simulations to arrive at this regime map, but in figure~\ref{fig:Regime_Map} we only show a few representatives at the transition lines.

For Newtonian fluids, the previous studies have found that for $\mathcal{O}h > 0.03$, viscous stresses dominate over the surface tension, such that the Worthington jet does not break up into droplets \citep{san2011size, ghabache2014physics, walls2015jet}. In this work, at $\mathcal{J} = 0$, we have reproduced this transition $\mathcal{O}h$ number (see left axis in figure~\ref{fig:Regime_Map}). Increasing $\mathcal{J}$ has a similar effect on the jet breakup as it manifests itself as increased apparent viscosity of the liquid. Consequently, even when $\mathcal{O}h \to 0$ the capillary waves get severely damped for $\mathcal{J} > 0.3$, and no droplets are formed. The blue area in figure~\ref{fig:Regime_Map} highlights the region in which a Worthington jet forms and disintegrate. 

The gray area in figure~\ref{fig:Regime_Map} shows an intermediate regime in which the jet forms  and crosses the free surface line $\left(\mathcal{Z} = 0\right)$ but does not breakup. This transition, for $\mathcal{J} = 0$, occurs at $\mathcal{O}h \approx 10^{-1}$. For non-zero $\mathcal{J}$, the transition occurs at smaller values of $\mathcal{O}h$  as the jet (if it forms) has less kinetic energy and cannot cross the $\mathcal{Z} = 0$ line. If the jet does not form (beyond the grey area), the collapse of the cavity results in a crater (for $\mathcal{J} \ne 0$). 

As discussed in \S~\ref{Sec::CapillaryWaves}, if the surface tension stresses are high enough, the whole cavity yields. Otherwise, for large values of the yield stress $\left(\mathcal{J} \sim \mathcal{O}\left(1\right)\right)$, the plug region attached to the bottom of the cavity never yields. As a result, the bottom of the cavity does not move, i.e., $\mathcal{H}_f = \mathcal{H}_i$. This transition from a fully yielded cavity to the cavity with an unyielded bottom is highlighted in figure~\ref{fig:Regime_Map} with the red line.

\section{Conclusions}\label{Sec::Conclusion}
In this work, we have studied the capillary driven process of bursting bubbles in a viscoplastic medium. Like in Newtonian fluids, flow begins when the rim, which connects the bubble cavity to the free surface. Consequently, the fluid is yielded, and a train of capillary waves is generated. The yield stress significantly affects the flow structure inside the pool by making plug regions. The higher the value of the yield stress, the larger the deviation from the Newtonian counterpart is. 

Following the analyses of \citet{deike2018dynamics} and \citet{gordillo2019capillary}, we provided information on the dynamics of the capillary waves as they travel down the bubble cavity. In liquids with low yield stresses, the cavity collapse leads to a Worthington jet that might break up into drops by a Rayleigh-Plateau instability. However, for liquids with a large yield stress, the jet vanishes. The energy budgets analysis gives insight into the dynamics by showing how the initial surface energy is dissipated. Eventually, in contrast to the Newtonian fluids, where the final state is always a flat film, bubble bursting in viscoplastic medium results in final crater shapes with high residual surface energy. We analysed the geometry of these shapes as a function of the governing control parameters, namely the Ohnesorge and the plastocapillary numbers. Lastly, we use the same numbers to categorise the four different regimes in viscoplastic bubble bursting (see the phase diagram in figure \ref{fig:Regime_Map}).

Our study has direct applications in a range of industrial operations, where bubbles are present at the surface of a yield stress fluids. The focus of this work is to compare the bursting bubble process in a yield stress fluid to that in a Newtonian fluid, without the initial shape effects. However, once the exact shape of the bubble at the free surface is known, either from experiments or theory, one can calculate the resulting flow and compare them to the present study. Moreover, the current results could be useful in analysing some geophysical flows, such as those in volcanic eruptions. In a broader perspective, the work presents a system in which surface tension and yield stress are the main factors. Such a system is of fundamental interest in design and manufacturing at small scales when capillary action is competing with the yield stress, e.g., in 3D printing and coatings polymeric fluids \citep{rauzan2018particle, nelson2019designing, jalaal2019laser, jalaal2021spreading}.\\

\noindent{\bf  Supplementary data\bf{.}} \label{SupMat} Supplementary material and movies are available at xxxx \\

\noindent{\bf Acknowledgements\bf{.}} We would like to thank Andrea Prosperetti, Arup Kumar Das, and Stéphaze Zaleski for insightful discussions about the Newtonian limit of bursting bubble process. We also want to thank Uddalok Sen, Rodrigo Ezeta, and Carola Seyfert for comments on the manuscript. This work was carried out on the national e-infrastructure of SURFsara, a subsidiary of SURF cooperation, the collaborative ICT organization for Dutch education and research.\\

\noindent{\bf Funding\bf{.}} The authors acknowledge the ERC Advanced Grant No. 740479-DDD.\\

\noindent{\bf Declaration of Interests\bf{.}} The  authors report no conflict of interest. \\

\noindent{\bf  Author ORCID\bf{.}} V. Sanjay, \href{https://orcid.org/0000-0002-4293-6099}{https://orcid.org/0000-0002-4293-6099}; D. Lohse, \href{https://orcid.org/0000-0003-4138-2255}{https://orcid.org/0000-0003-4138-2255}; M. Jalaal \href{https://orcid.org/0000-0002-5654-8505}{https://orcid.org/0000-0002-5654-8505};\\

\appendix
\section{Governing equations}\label{App::GoverningEquations}
In this appendix, we describe the governing equations that describe the process of bursting bubbles in a viscoplastic medium. For an incompressible liquid, the continuity and momentum equations read  
\begin{align}\label{Eqn::NS_liq}
\nabla\boldsymbol{\cdot}\boldsymbol{u} &= 0,\\ 
\label{Eqn::NS_liq2}
\rho_l\left(\frac{\partial\boldsymbol{u}}{\partial t} + \nabla\boldsymbol{\cdot}\left(\boldsymbol{uu}\right)\right) &= -\nabla p + \nabla\boldsymbol{\cdot}\boldsymbol{\tau} + \rho_l\boldsymbol{g},
\end{align}

\noindent where $\boldsymbol{u}$ is the velocity vector, $\rho_l$ is the density of the liquid, $p$ is the pressure field, $\boldsymbol{\tau}$ is the stress tensor in liquid, and $\boldsymbol{g}$ is the acceleration due to gravity.
We model the viscoplastic liquid medium as a non-Newtonian Bingham fluid with a yield stress, $\tau_y$. For such liquids, the constitutive equation are
\begin{equation}\label{Eqn::ConstitutiveEqns}
\begin{cases}
\boldsymbol{\mathcal{D}}=0\,\,&\,\,\|\boldsymbol{\tau}\| < \tau_y\\
\boldsymbol{\tau} =\left( \frac{\tau_y}{2 \|\boldsymbol{\mathcal{D}} \|} + \mu_l \right)\,\,&\,\,\|\boldsymbol{\tau}\| \geq \tau_y
\end{cases}
\end{equation} 
In the equation above, $\boldsymbol{\mathcal{D}} = \left(\nabla\boldsymbol{u} + \left(\nabla\boldsymbol{u}\right)^{\text{T}}\right)/2$ is the deformation tensor and $\mu_l$ the constant viscosity in the Bingham model. 
We adopt a regularised revision of equation~(\ref{Eqn::ConstitutiveEqns}) in our numerical simulations, given by:
\begin{equation}\label{Eqn::NS_liq_Stress}
\boldsymbol{\tau} = 2\,\mathrm{min}\left(\frac{\tau_y}{2\|\boldsymbol{\mathcal{D}}\|} + \mu_l , \mu_{max}\right) \boldsymbol{\mathcal{D}}
\end{equation}
In equation~(\ref{Eqn::NS_liq_Stress}), $\frac{\tau_y}{2\|\boldsymbol{\mathcal{D}}\|} + \mu_l$ is basically the apparent viscosity $\left(\mu_{\text{eff}}\right)$ of the liquid and $\mu_{\text{max}}$ is the \lq\lq large\rq\rq regularisation viscosity, such that $\mu_{\text{eff}} \gets \text{min}\left(\mu_{\text{eff}}, \mu_{\text{max}}\right)$. 

The same sets of mass and momentum conservation equations~\ref{Eqn::NS_liq} --~\ref{Eqn::NS_liq2} are also solved for the gas phase, but now with constant density and viscosity. We use the inertia-capillary velocity $\left(V_\gamma\right)$ and inertial-capillary time $t_\gamma$, and the capillary stress $\tau_\gamma$ defined as
\begin{align}\label{Eqn::Scales1}
	V_\gamma = \sqrt{\frac{\gamma}{\rho_lR_0}},&\,\,\,t_\gamma = \frac{R_0}{V_\gamma} = \sqrt{\frac{\rho_lR_0^3}{\gamma}},\\
	\label{Eqn::Scales2}
	&\tau_\gamma = \frac{\gamma}{R_0},
\end{align}

\noindent to non-dimensionalize above governing equations to find equations~(\ref{Eqn::FinalFormLiqContinuity}) to~(\ref{Eqn::DimensionlessNumbers}).

\section{The Newtonian limit}\label{App::Validation}
\begin{figure}
	\centerline{\includegraphics[width=\linewidth]{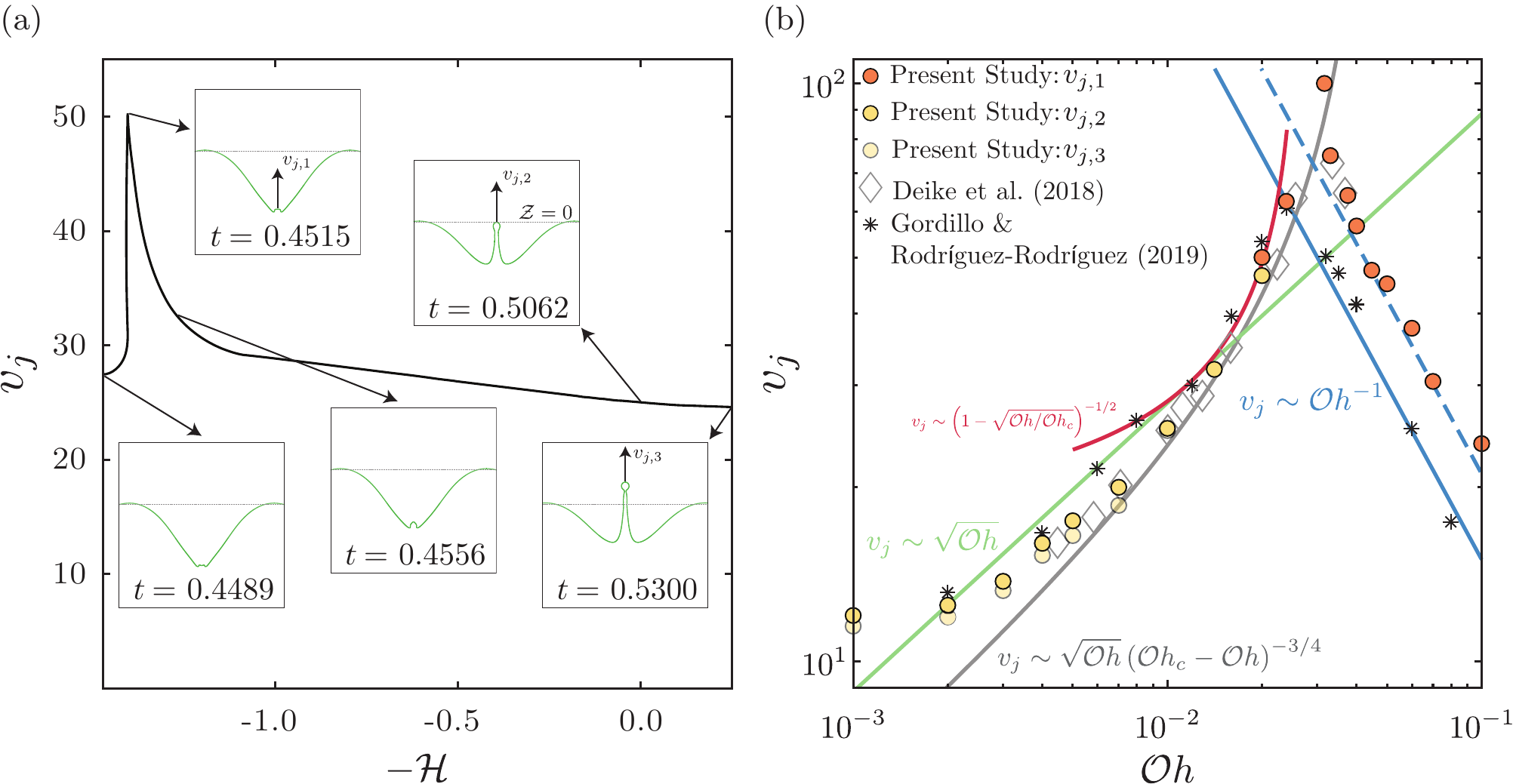}}
	\caption{Characterization of the Worthington jet's velocity formed as a result of the bursting bubble process in Newtonian liquids: (a) Variation of the jet's velocity as it travels through different axial locations ($\mathcal{O}h = 10^{-2}$). The inset shows the shape of this jet at different time. The grey dotted line represents the free surface, $\mathcal{Z} = 0$. (b) Comparison of the jet's velocity with the data and scaling laws available in the literature for the range of Ohnesorge numbers used in this study. Note that the scaling law in solid grey line comes from \citet{deike2018dynamics}, whereas the other lines are from \citet{gordillo2019capillary} as noted in the figure. }
	\label{fig:NewtonianJetVelocity}
\end{figure}
\begin{figure}
	\centerline{\includegraphics[width=\linewidth]{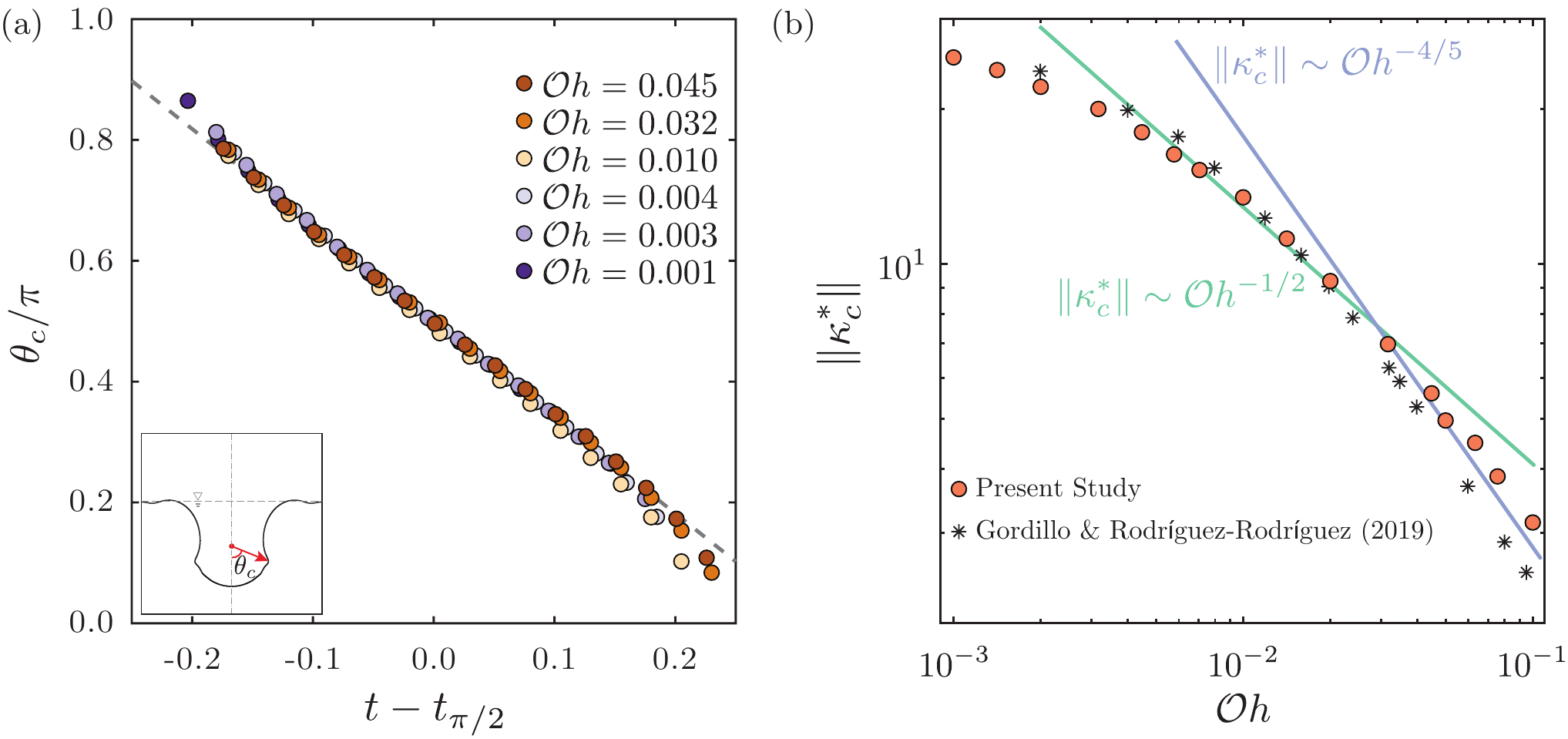}}
	\caption{(a) Variation of the location $\theta_c$ of the strongest capillary wave with time. The gray dotted line denotes $\theta_c - \theta_i \sim -V_\gamma t$ as described by \citet{gordillo2019capillary}. (b) Variation of the strength $\|\kappa_c^*\|$ at $\theta_c = \pi/2$ with the $\mathcal{O}h$ numbers. The scaling laws are taken from \cite{gordillo2019capillary}.}
	\label{fig:CapillaryWavesComparision}
\end{figure}
One of the essential and widely studied features of the bursting bubble process in a Newtonian liquid is the resulting Worthington jet's velocity. This jet is formed because of the strong flow-focusing caused by the capillary waves at the bubble cavity's bottom. In general, this process is very fast, as shown in figure~\ref{fig:NewtonianJetVelocity}(a). Over a small time span of $\approx 0.1t_\gamma$ (see insets of figure~\ref{fig:NewtonianJetVelocity}), the jet traverses a distance of $\approx 1.5R_0$. Moreover, the inception of this jet is characterized by velocities as high as $50V_\gamma$.  This jet's flow is also associated with high viscous dissipation (because of the high strain rates resulting from such high velocities). As a result of these two processes, there is a distinct maximum at the instant of jet inception  $\left(v_{j,1}\right)$. This velocity could be difficult to calculate, especially at low $\mathcal{O}h$ numbers because of high-frequency capillary waves. Numerically, it is easiest to calculate the velocity of the jet as it crosses the free surface, $\mathcal{Z} = 0$ (grey dotted line in figure~\ref{fig:NewtonianJetVelocity}(a)). However, in experiments, it is easier to calculate the velocity of the first droplet that forms as a result of the jet breakup. In the inset of figure~\ref{fig:NewtonianJetVelocity}(a), the instant immediately before jet breakup into a droplet gives a velocity of $v_{j,3}$. As a result, in the literature, different authors have reported different jet velocities. We have decided to plot all three velocities (wherever applicable) in figure~\ref{fig:NewtonianJetVelocity}(b) along with the scaling laws proposed by \citet{deike2018dynamics} (grey line) and \citet{gordillo2019capillary} (green, red, and blue lines). Our results agree well with the previously published works, which have been extensively validated with experimental data. Note that the differences between our data points and those of \citet{gordillo2019capillary} also arise because of a slight difference in Bond numbers for the two studies ($\mathcal{B}o = 5\times 10^{-2}$ in \citet{gordillo2019capillary} as compared to $\mathcal{B}o = 10^{-3}$ in \cite{deike2018dynamics} and in the present work). This disagreement is higher for high $\mathcal{O}h$ numbers. Furthermore, as pointed out by \citet{deike2018dynamics}, at lower $\mathcal{B}o$, the maxima in the $v_j - \mathcal{O}h$ plot shifts to the right with higher velocities, a feature which is distinctly captured by figure~\ref{fig:NewtonianJetVelocity}(b).

Figure~\ref{fig:CapillaryWavesComparision}(a) shows the temporal evolution of the angular trajectory of the strongest capillary wave as it travels down the bubble cavity. As predicted by \citet{gordillo2019capillary} and shown experimentally by \citet{krishnan2017scaling}, this wave travels at a constant angular velocity, implying $\theta_c - \theta_i \sim -V_\gamma t$ (gray dotted line in figure~\ref{fig:CapillaryWavesComparision}(a)). Furthermore, we also compare the strength of this wave with those predicted by the scaling laws given in \citet{gordillo2019capillary} and found good agreement (figure~\ref{fig:CapillaryWavesComparision}(b)).  

\section{The effect of $\mathcal{O}h$}\label{App::OhsVariation}
 \begin{figure}
	\centerline{\includegraphics[width=\linewidth]{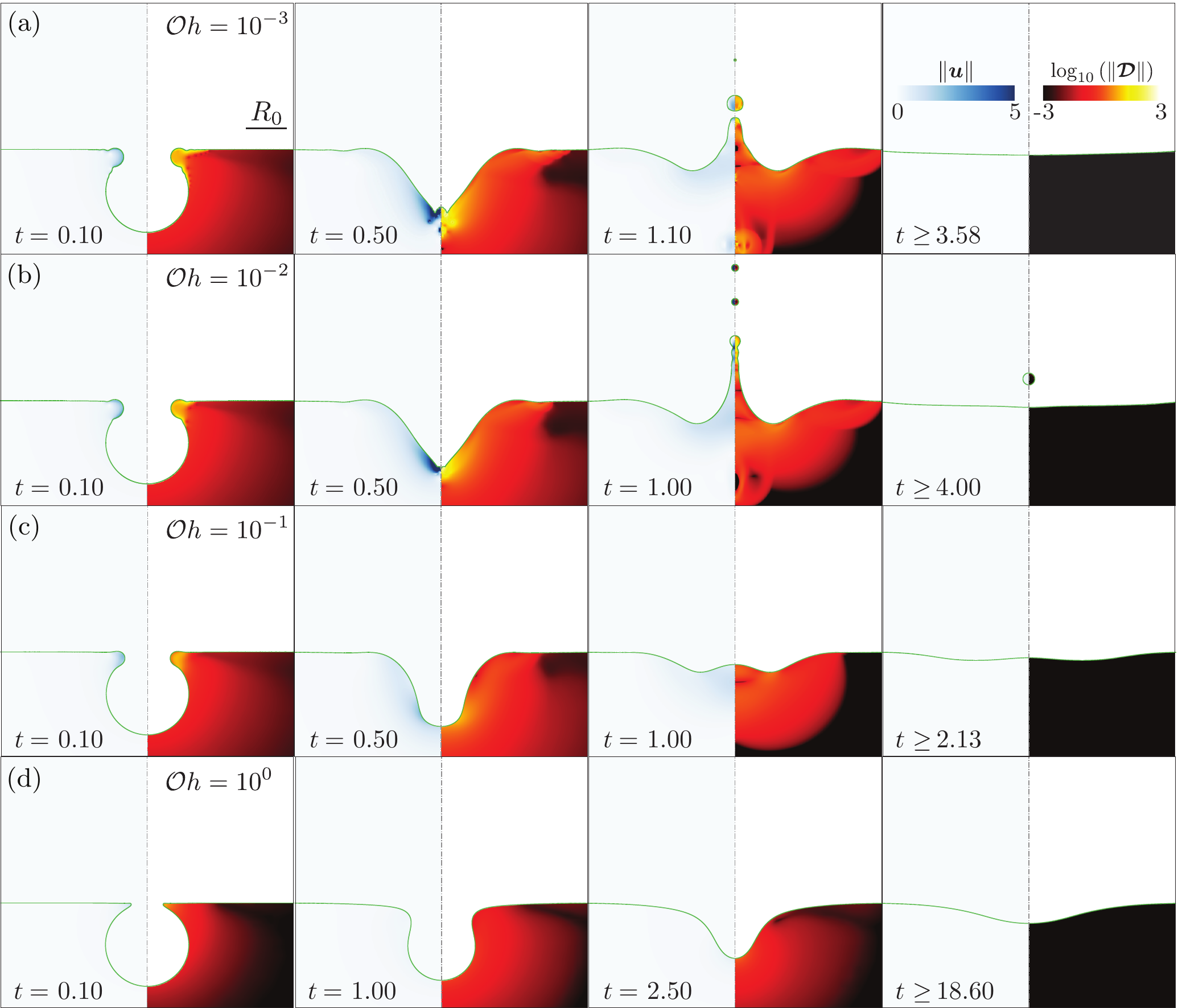}}
	\caption{Bursting bubble dynamics for different Ohnesorge numbers: (a) $\mathcal{O}h = 10^{-3}$, (b) $\mathcal{O}h = 10^{-2}$, (c) $\mathcal{O}h = 10^{-1}$, and (d) $\mathcal{O}h = 10^{0}$. In the background, the left part of each panel shows the magnitude of the velocity field and the right part shows the deformation tensor's magnitude on a $\log_{10}$ scale. For all the cases in this figure, $\mathcal{J} = 0.1$. Videos (S2 and S11 - S13) are available in the supplementary material.}
	\label{fig:Oh_Variation}
\end{figure}
This appendix describes the dynamics of bursting bubbles and the qualitative effects of varying Ohnesorge number $\mathcal{O}h$ at given plastocapillary number of $\mathcal{J}=0.1$. Figure~\ref{fig:Oh_Variation} illustrates four representative cases for this purpose. As noted in the text and several previous studies \citep{duchemin2002jet, deike2018dynamics, gordillo2019capillary}, the initial retraction of the rim forms a train of capillary waves. For low Ohnesorge numbers, e.g., $\mathcal{O}h = 10^{-3}$ in figure~\ref{fig:Oh_Variation}(a), the viscous dissipation is very small, and as a result, most of these capillary waves converge at the bottom of the cavity and result in vigorous surface undulations at the cavity's bottom (figure~\ref{fig:Oh_Variation}(a): $ t = 0.1 - 0.50$). These waves result in a thick Worthington jet. As the $\mathcal{O}h$ increases ($\mathcal{O}h = 10^{-2}$ in figure~\ref{fig:Oh_Variation}(b)), viscous dissipation damps the high-frequency capillary waves and improves the flow focusing at the cavity's bottom, leading to thinner and faster jets. This process is similar to what has been reported in the literature for Newtonian liquids \citep{duchemin2002jet, ghabache2014physics, deike2018dynamics}. Note that figure~\ref{fig:Oh_Variation}(b) is the same as figure~\ref{fig:J_Variation}(b) and has been presented again for completeness. 

Larger $\mathcal{O}h$ numbers ($10^{-1}\,\&\,10^{0}$ in figures~\ref{fig:Oh_Variation}(c) and~\ref{fig:Oh_Variation}(c), respectively), result in a longer flow time scales.	Nonetheless, at low $\mathcal{J}$ numbers (such as $0.1$ in figure~\ref{fig:Oh_Variation}), the entire cavity still yields and the center gently approaches the free surface at $\mathcal{Z} = 0$ (figure~\ref{fig:Oh_Variation}(b) and~\ref{fig:Oh_Variation}(c): first three columns). Most of the initial surface energy is lost as viscous dissipation (both $E_d^{\mathcal{O}h}\,\&\, E_d^{\mathcal{J}}$) and the flow stops as the internal stresses in the fluid falls below the yield stress (figure~\ref{fig:Oh_Variation}(b) and~\ref{fig:Oh_Variation}(c): last column).

\section{Energy budget calculations}\label{App::EnergyBudgetApp}
Here, we describe the formulation used to evaluate the different energy transfer modes discussed in \S~\ref{Sec::Energy}. A similar approach was used by \citet{wildeman2016spreading} and by \citet{ramirez2020lifting} to evaluate the energy budget for impacting droplets and colliding droplets, respectively. In this work, we have extended the methodology to yield-stress liquids. The kinetic and the surface energy of the liquid are given by
\begin{align}
	\label{Eqn::Ek}
	E_k &= \frac{1}{2}\int_{\Omega_p}\|\boldsymbol{u}\|^2\,\mathrm{d}\Omega_p,\\
	\label{Eqn::Es}
	E_s &= \int_{\Gamma_p}\,\mathrm{d}\Gamma_p,
\end{align}

\noindent where the energies are normalized by the surface energy $\gamma \, R_0^2$. The integrals are evaluated over the volume $\left(\Omega_p\right)$ and the surface $\left(\Gamma_p\right)$ of the biggest liquid continuum in the domain, disregarding the drops (which are included in the energy budget in a different way described below). The state of liquid pool with a flat free-surface is taken as the reference to calculate $E_s$. 

Extending the Newtonian fluid's formulation in \citet[p.~50-51]{landau2013course}, the total dissipation in our system can be calculated as

\begin{equation}\label{Eqn::Ed}
	E_d = 2\int_t\left(\int_{\Omega_p}\left(\mathcal{O}h + \frac{\mathcal{J}}{2\|\boldsymbol{\mathcal{D}}\|}\right)\|\boldsymbol{\mathcal{D}}\|^2\,\mathrm{d}\Omega_p\right)\mathrm{d}t.
\end{equation}

\noindent Note that by writing the equation in this form, we assume that the yield stress contributes to the energy dissipation only through an increase in the effective viscosity (see Appendix~\ref{App::GoverningEquations}). In order to isolate the effects of the viscosity and yield-stress associated viscosity, we can rewrite equation~(\ref{Eqn::Ed}) as $E_d = E_d^{\mathcal{O}h} + E_d^\mathcal{J}$, where
\begin{align}
	\label{Eqn::EdOh}
	E_d^{\mathcal{O}h} = 2\, \mathcal{O}h&\int_t\left(\int_{\Omega_p}\|\boldsymbol{\mathcal{D}}\|^2\,\mathrm{d}\Omega_p\right)\mathrm{d}t,\\
	\label{Eqn::EdJ}
	E_d^\mathcal{J} = \mathcal{J}&\int_t\left(\int_{\Omega_p}\|\boldsymbol{\mathcal{D}}\|\,\mathrm{d}\Omega_p\right)\mathrm{d}t.
\end{align}

We present together all other forms of energy as

\begin{equation}\label{Eqn::Em}
	E_m = E_k^\text{Drops} + E_s^\text{Drops} + E_d^\text{Drops} + \int_{\Omega_p+\Omega_d}\mathcal{B}o\,\mathcal{Z}\,\mathrm{d}\left(\Omega_p+\Omega_d\right)  + E_g. 
\end{equation}

 In equation~(\ref{Eqn::Em}), the first two terms, $E_k^\text{Drops}\,\&\,E_s^\text{Drops}$ denote the kinetic and the surface energies of the ejected drops, respectively. The third term, $E_d^\text{Drops}$, is the sum of the effective dissipation inside the drop. Note that all these three terms are evaluated like equations~\ref{Eqn::Ek} to~\ref{Eqn::EdJ} with one difference that the volume and surface integrals are performed over the drops ($\Omega_d$ and $\Gamma_d$, respectively), instead of over the pool ($\Omega_p$ and $\Gamma_p$, respectively). The next term evaluates the gravitational potential energy for the liquid (both the pool and the drops). As $\mathcal{B}o \to 0$, this term is insignificant. Lastly, $E_g$ denotes the sum of energies stored in the gas medium and viscous dissipation due to velocity gradients inside it:
\begin{equation}\label{Eqn::Eg}
	E_g = \rho_r\int_{\Omega_g}\left(\frac{\|\boldsymbol{v}\|^2}{2} + \mathcal{B}o\,\mathcal{Z}\right)\mathrm{d}\Omega_g + 2\mu_g\mathcal{O}h\int_t\left(\int_{\Omega_g}\|\boldsymbol{\mathcal{D}}\|^2\mathrm{d}\Omega_g\right)\mathrm{d}t.
\end{equation}

\noindent$E_m$ (equation~(\ref{Eqn::Em})) is only significant when the resultant Worthington jet leads to the formation of droplets (figure~\ref{fig:EnergyBudget}(c)). 

\section{Code availability \& choosing numerical parameters}\label{App::codes}
For our calculations, we use the free software program Basilisk C \citep{basiliskPopinet, popinet2015quadtree}. To ensure reproducibility, the codes used in the present article are permanently available at \citet{basiliskVatsal}. Furthermore, \S~\ref{Sec::ProblemDescription} contains major computational choices and parameters employed in the current study. In this appendix, we provide further details and reasons for selecting the critical parameters in light of the regularisation method. 

\subsection{Viscous regularisation parameter}\label{App::OhMax}
\begin{figure}
	\centerline{\includegraphics[width=\linewidth]{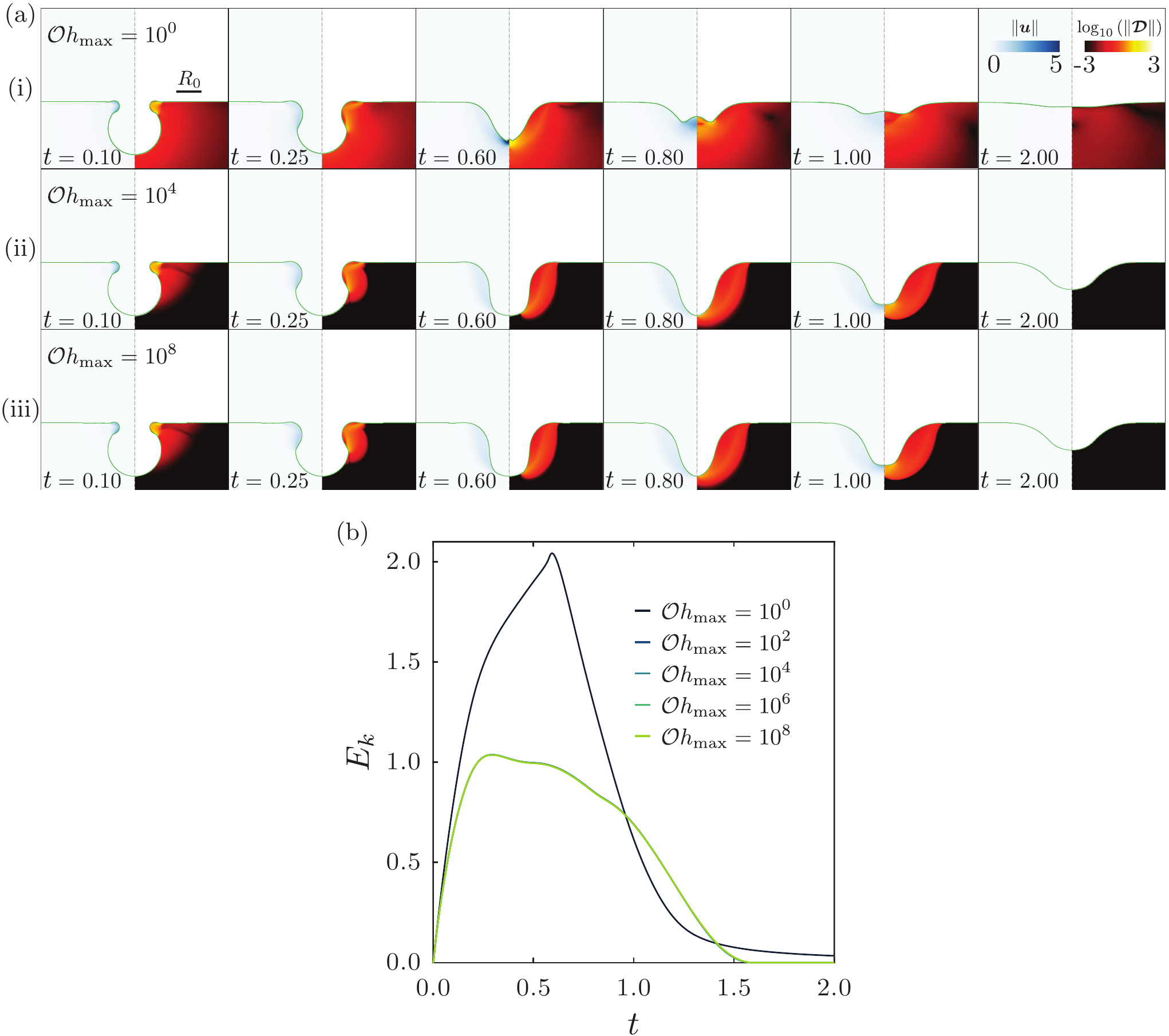}}
	\caption{Sensitivity to viscous regularisation parameter $\mathcal{O}h_{\text{max}}$: Temporal evolution of the bubble cavity for (a) $\mathcal{O}h_{\text{max}} =$ (i) $10^0$, (ii) $10^4$, and (iii) $10^8$, and (b) Kinetic energy evolution in time. The results show negligible differences for $\mathcal{O}h_{\text{max}} > 10^2$.}
	\label{fig:OhMax}
\end{figure}
To verify that the macroscopic flow features were independent of the regularisation parameter, we conducted simulations for different $\mathcal{O}h_\text{max}$. We show a representative case for this test in figure~\ref{fig:OhMax}. Using a small value of $\mathcal{O}h_\text{max}$, such as $10^0$ in figure~\ref{fig:OhMax}(a - i), the process resembles a case with increased effective viscosity. However, at higher values of $\mathcal{O}h_\text{max}$ ($10^4$ for figure~\ref{fig:OhMax}(a - ii) or $10^8$ for figure~\ref{fig:OhMax}(a - iii)), the process is independent of viscous regularisation parameter. To ensure that the flow is captured precisely, the liquid kinetic energy is tracked over time as well (figure~\ref{fig:OhMax}(b)). For $\mathcal{O}h_{\text{max}} > 10^2$, there is negligible differences between the cases. 

Comparing the values of $\|\boldsymbol{\mathcal{D}}\|$, it is shown that the flow patterns are unchanged for a given large values of $\mathcal{O}h_{\text{max}}$ and once could clearly distinguish a sharp transition between a weakly deformed region and a strongly deformed one. That being said, we would like to mention that the identification of the yield surface is obscured, when regularized constitutive models are used (\textit{cf.} \cite{frigaard2005usage}). Nevertheless, irrespective of such details, the map of the second invariant of the deformation rate tensor provides important information on flow patterns inside viscoplastic medium.

\subsection{Stoppage time}\label{App::StoppageTime}
\begin{figure}
	\centerline{\includegraphics[width=\linewidth]{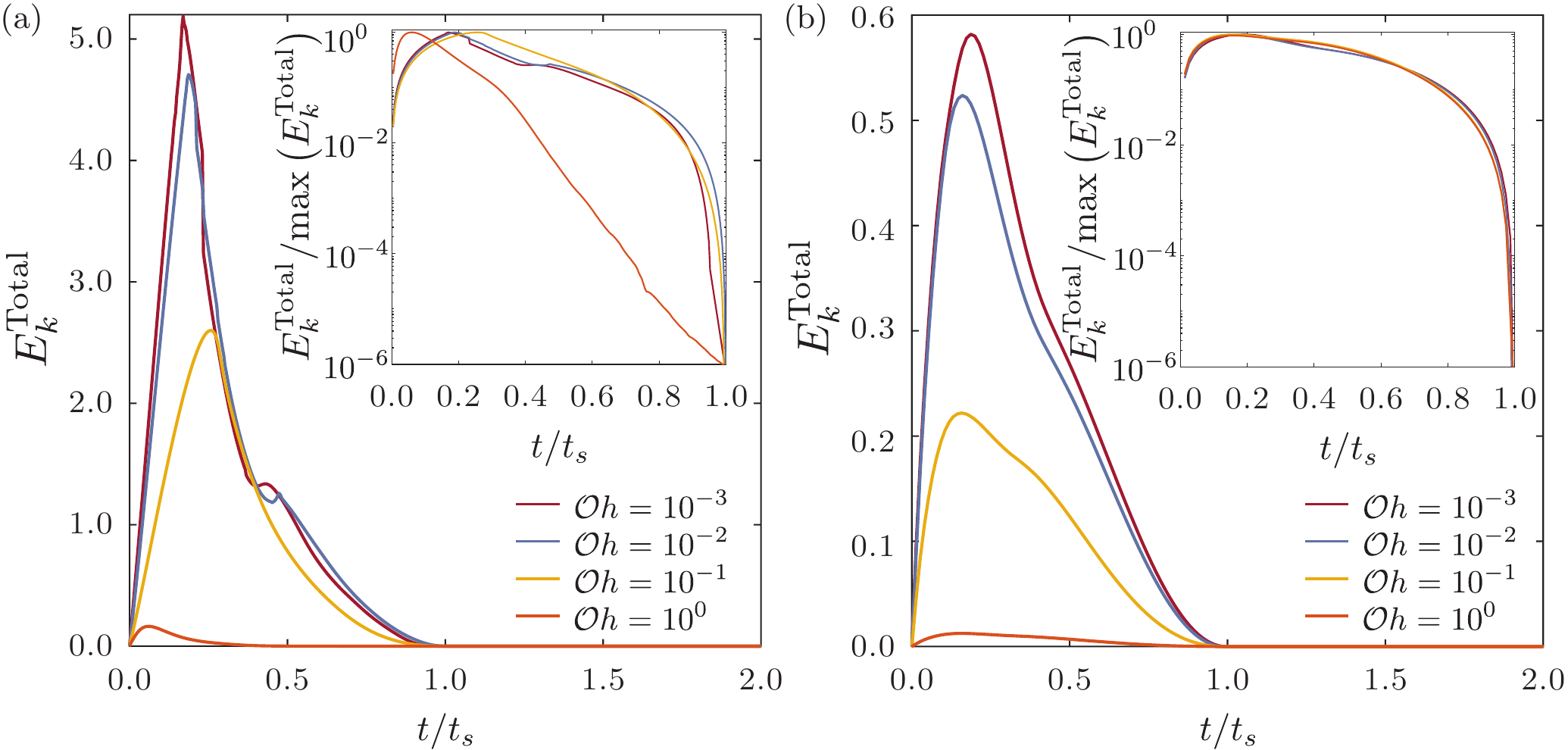}}
	\caption{Selection of stoppage time: Variation of the kinetic energy of the liquid, $E_k$ (see equation~(\ref{Eqn::E_kTotal})) with time for eight representative cases: (a) $\mathcal{J} = 0.1$ \& $\mathcal{O}h = 10^{-3} - 10^0$ and (b) $\mathcal{J} = 1.0$  \& $\mathcal{O}h = 10^{-3} - 10^0$. The inset of each figure shows the kinetic energy normalized by the maximum kinetic energy on a semi-log scale. We define a finite stoppage time, $t = t_s$ beyond which the flow is too slow to cause any macroscopic changes in the time-scales that we want to study.}
	\label{fig:StoppageTime}
\end{figure}

Another important consequence of yield stress is a finite stoppage time. The flow in a liquid will stop if the stress falls below the yield stress. This implies that $\|\boldsymbol{\mathcal{D}}\|$ should vanish. Since we use a regularisation method, the flow in our simulations never truly stops. Hence, we consider a cut-off kinetic energy of $10^{-6}\,\times\,\text{max}(E_k^\text{Total})$, where
\begin{equation}\label{Eqn::E_kTotal}
	E_k^\text{Total} = 0.5\int_\Omega\left(\|\boldsymbol{u}\|^2 + \rho_r\|\boldsymbol{v}\|^2\right)\mathrm{d}\Omega
\end{equation}
is the total kinetic energy of the system. We stop the calculations when the total kinetic energy of the system is below the cut-off. To verify the sensitivity of this cut-off, we ran a number of simulations till $t = 2t_s$ (figure~\ref{fig:StoppageTime}). Clearly, independent of the $\mathcal{O}h$ number, the flow becomes asymptotically stationary beyond $t = t_s$. A sharp decrease in the total kinetic energy can be observed in the semi-log inset plots of Figures~\ref{fig:StoppageTime}(a) and~\ref{fig:StoppageTime}(b) as stoppage time is approached. Note that, this analysis only provides an estimation for the stoppage time, $t_s$ and a more comprehensive study is required to find the exact values of $t_s$. Nonetheless, as clear from our results like those shown in  figures~\ref{fig:J_Variation} and~\ref{fig:StoppageTime}, beyond this time, the flow dynamics are too slow for any macroscopic change in the location or the strength of capillary waves, or the shape of the final cavity.

\subsection{Initial rim curvature effects}\label{App::FilletCurvature}
\begin{figure}
	\centerline{\includegraphics[width=\linewidth]{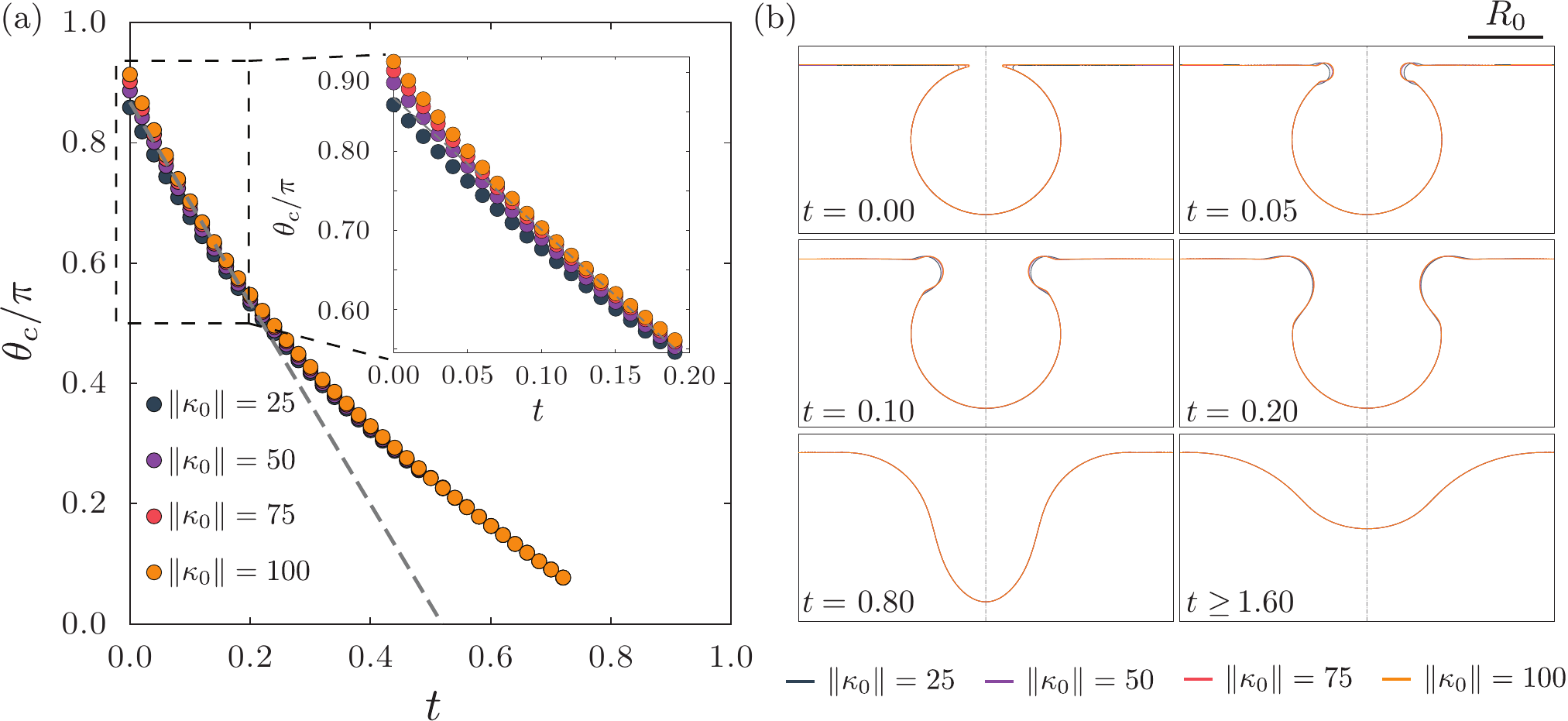}}
	\caption{Sensitivity to initial rim curvature: (a) Temporal evolution of the location of the strongest capillary wave $\theta_c$, and (b) Influence on the overall process of cavity collapse. Beyond, $t = 0.2$, there is negligible difference between the interfaces as the effect of initial condition vanishes. Inset in (a) zooms into the initial stages of the process where influence of $\|\kappa_0\|$ is apparent.}
	\label{fig:kappa0}
\end{figure}
As mentioned in \S~\ref{Sec::Initial}, in our initial shape, we introduced a rim with a finite curvature $\kappa_0$ that connects the bubble to the free surface. The initial location of the cavity-free surface intersection $\theta_i$ also changes (inset of figure~\ref{fig:kappa0}(a)), depending on this regularised curvature. In this appendix, we will show how this initial condition affects the bubble bursting process. For all the cases presented in this work, we have used $\kappa_0 = 100$. For Newtonian cases, \citet{deike2018dynamics} has carried out a more extensive sensitivity test to ascertain the importance of initial cavity shape on the process. For $\mathcal{J} = 0$, our results support their findings that the size of the initial hole around the axis ($\mathcal{R} = 0$) is crucial and can manifest into changes in the jet velocity. Furthermore, $\kappa_0$ controls the first capillary waves that appear as the bubble cavity collapses (\S~\ref{Sec::CapillaryWaves}). Figures~\ref{fig:NewtonianJetVelocity} and~\ref{fig:CapillaryWavesComparision} show that our results agree with \cite{deike2018dynamics, gordillo2019capillary} which have been validated with experiments.

For higher $\mathcal{J}$ values, it is essential that $\|\kappa_0\| > \mathcal{J}$ for flow initiation. We have restricted our study to $\mathcal{J} = 64$ where the flow is confined in the region of this high curvature (see \S~\ref{Sec::EquilibriumStates}). Furthermore, figure~\ref{fig:kappa0} contains one representative case where we show the influence of $\kappa_0$ on the temporal evolution of the strongest capillary wave as it travels down the bubble cavity (figure~\ref{fig:kappa0}(a)) and also on the interface deformations (figure~\ref{fig:kappa0}(b)). As shown in these curves the difference in the results is negligible when $\|\kappa_0\| > 75$.

\bibliographystyle{jfm}
\bibliography{BurstingBubble}
\end{document}